\documentclass[preprint,12pt]{elsarticle}
\usepackage[a4paper, total={6.5in, 10in}]{geometry}
\usepackage{lineno}
\usepackage[export]{adjustbox}
\journal{\texttt{arXiv}}
 
\RequirePackage{fontenc,inputenc,subcaption,graphicx,sectsty,titlesec, caption, fancyhdr}
\RequirePackage{amsthm,amsmath,amssymb,nccmath}
\RequirePackage{paralist,multirow,arydshln}%algorithm, 
\RequirePackage[ruled, vlined]{algorithm2e}
\RequirePackage{float}

\DeclareMathOperator*{\argmin}{arg\,min}

\newcommand{\trans}{^{T}} % transpose
\newcommand{\del}[2]{\frac{\partial #1}{\partial #2}} % partial-differentiation
\newcommand\scalemath[2]{\scalebox{#1}{\mbox{\ensuremath{\displaystyle #2}}}}

\def\balpha{\mbox{\boldmath$\medmath{\alpha}$}}
\def\bbeta{\mbox{\boldmath$\medmath{\beta}$}}
\def\bgamma{\mbox{\boldmath$\medmath{\gamma}$}}
\def\bdelta{\mbox{\boldmath$\medmath{\delta}$}}

\def\bleta{\mbox{\boldmath$\medmath{\eta}$}}

\newtheorem{mythe}{Theorem}

\newtheorem{mylem}{Lemma}
\newtheorem{myrem}{Remark}

\begin{document}

\begin{frontmatter}

\title{Spatial Tweedie exponential dispersion models}
%\tnotetext[mytitlenote]{Fully documented templates are available in the elsarticle package on \href{http://www.ctan.org/tex-archive/macros/latex/contrib/elsarticle}{CTAN}.}

%% Group authors per affiliation:
\author{Aritra Halder\corref{mycorrespondingauthor}}
\cortext[mycorrespondingauthor]{Corresponding author}
\ead{aritra.halder@uconn.edu}
\author{Shariq Mohammed}
\author{Kun Chen}
%\author[ternaryaddress]{Brien Aronov}
\author{Dipak K. Dey}
%\fntext[myfootnote]{Since 1880.}

%% or include affiliations in footnotes:
% \ead[url]{www.elsevier.com}
%\author[mysecondaryaddress]{Global Customer Service}
%\address{University of Connecticut, Storrs, CT - 06269, USA.}

\address{Department of Statistics, University of Connecticut, Storrs, CT - 06269, USA.}
%\address[ternaryaddress]{Travelers Insurance, 1 Tower Square, Hartford, CT--06183, USA}

\begin{abstract} 
	
	\vspace*{.25in}
	
	\small
	This paper proposes a general modeling framework that allows for uncertainty quantification at the individual covariate level and spatial referencing, operating withing a double generalized linear model (DGLM). DGLMs provide a general modeling framework allowing dispersion to depend in a link-linear fashion on chosen covariates. We focus on working with Tweedie exponential dispersion models while considering DGLMs, the reason being their recent wide-spread use for modeling mixed response types. Adopting a regularization based approach, we suggest a class of flexible convex penalties derived from an un-directed graph that facilitates estimation of the unobserved spatial effect. Developments are concisely showcased by proposing a co-ordinate descent algorithm that jointly explains variation from covariates in mean and dispersion through estimation of respective model coefficients while estimating the unobserved spatial effect. Simulations performed show that proposed approach is superior to competitors like the ridge and un-penalized versions. Finally, a real data application is considered while modeling insurance losses arising from automobile collisions in the state of Connecticut, USA for the year 2008.

	\vspace*{.25in}

\end{abstract}


\end{frontmatter}

%\linenumbers

\section{Introduction}

Prolific geo-tagging of recorded data has allowed for exploration and development of rich frameworks for spatial modeling that permit quantification of the nature of dependence for a chosen response on a specified underlying spatial domain of reference. The type of referencing is broadly divided into (a) areal and, (b) point. Areal referencing involves use of county, zip-code, block or census-tract information, while point referencing relies on stereo-graphic projective co-ordinates, i.e. latitudes and longitudes coupled with projection on a sphere, $\mathbb{S}^2$. A projection on $\mathbb{S}^2$ is generally asymptomatic to the earth, with a purpose of providing subjective basis to the spatial analysis of chosen response. Focusing on areally referenced data generating processes, generally the response along with covariates is recorded at an individual level for multiple individuals residing at each areal co-ordinate. Consequently variation in response can be divided into two sources viz., {\em within} and {\em between} areal co-ordinates. An ideal model should jointly specify components that account and characterize both sources of variation. Explicitly, this translates to equal importance being given to modeling individual level covariates along with accounting for additional noise introduced by the variability across areal co-ordinates, when creating a modeling framework. In this paper we aspire to create such a modeling framework for a widely used and well celebrated family of probability distributions.

The exponential dispersion (ED) models \cite{jorgensen1997theory}, have provided a flexible modeling framework for a variety of data generating processes over the last century. Flexibility of such models lie in their ability to account for a variety of perturbations or deviations from the simple linear model framework viz., discreteness, over-dispersion, zero-inflation etc. A comprehensive framework is provided by the generalized linear models (GLMs) which incorporate these probability densities (\cite{nelder1972generalized}, \cite{mccullagh2018generalized}) within a modeling framework. With each mentioned perturbation, enrichment of the GLM framework proceeds by relaxing unnecessary restrictions, for example requirement of normality); introducing components with a purpose of specifying variation from a specific source, for example frailty models in survival analysis (see \cite{hougaard1995frailty}, \cite{banerjee2007bayesian} and \cite{ibrahim2014b}) or providing additional probability mass at chosen points in the support, for example zero-inflation (\cite{hall2000zero}, \cite{agarwal2002zero}, \cite{wang2015bayesian} and, \cite{zhang2013cplm}). They are typically specified using a {\em link function} that relates the natural parameter to a linear function of chosen covariates. In this paper we concentrate on a particular generalization of GLMs namely the double generalized linear models (DGLMs) \cite{pregibon1984} which do not require the dispersion to be fixed across observations. DGLMs have been studied in the context of modeling insurance losses and have proven to be effective, in allowing the investigator to examine the mean-variance relationship through separate models for mean and dispersion (\cite{jorgensen1994fitting}), \cite{smyth2002fitting}) under the assumption of a compound Poisson-gamma (CP-g) response. The CP-g density is a particular member of the ED family, in particular they belong to a more widely known class, the Tweedie ED models (\cite{tweedie1984index}). They feature a positive probability mass at zero along with continuous positive support. Tweedie CP-g GLMs have seen widespread application in a variety of fields including but not limited to modeling precipitation, economic donations, ecological biomass and insurance premiums (\cite{dunn2003precipitation}, \cite{smyth1996regression}, \cite{shono2008application}, \cite{foster2013poisson}, \cite{lauderdale2012compound}, \cite{dons2016indirect}, \cite{zhang2013likelihood}, \cite{yang2018insurance}). To mention other well studied members of Tweedie ED family: Gaussian, Poisson, Gamma and Inverse Gaussian. It provides an additional index parameter to distinguish between these members.

\begin{figure}[t]
	\centering
	\begin{subfigure}{.33\textwidth}
		\centering
		\includegraphics[width=1\linewidth , height=1\linewidth]{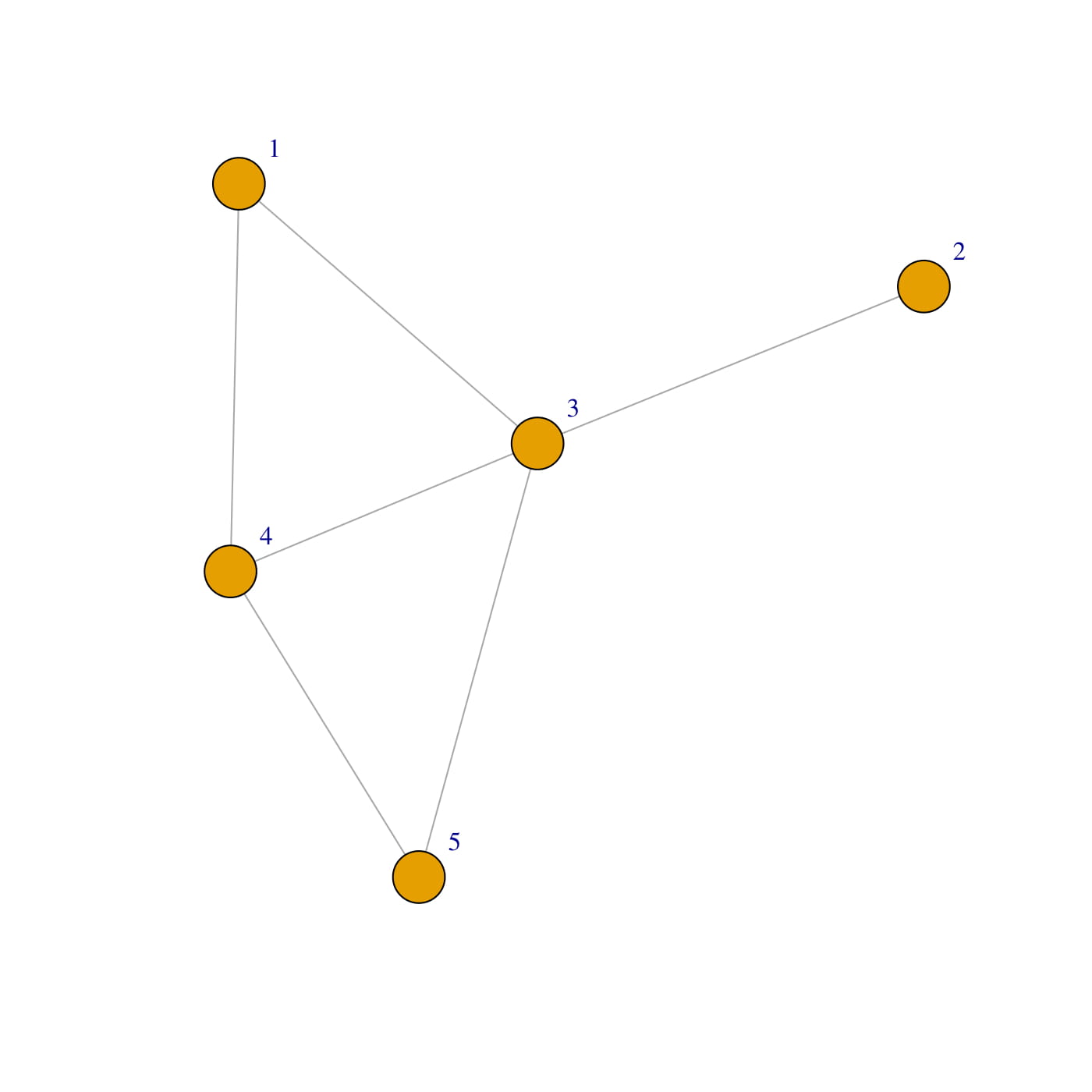}
		\caption{}
	\end{subfigure}%a
	\begin{subfigure}{.33\textwidth}
		\centering
		\includegraphics[width=1\linewidth , height=1\linewidth]{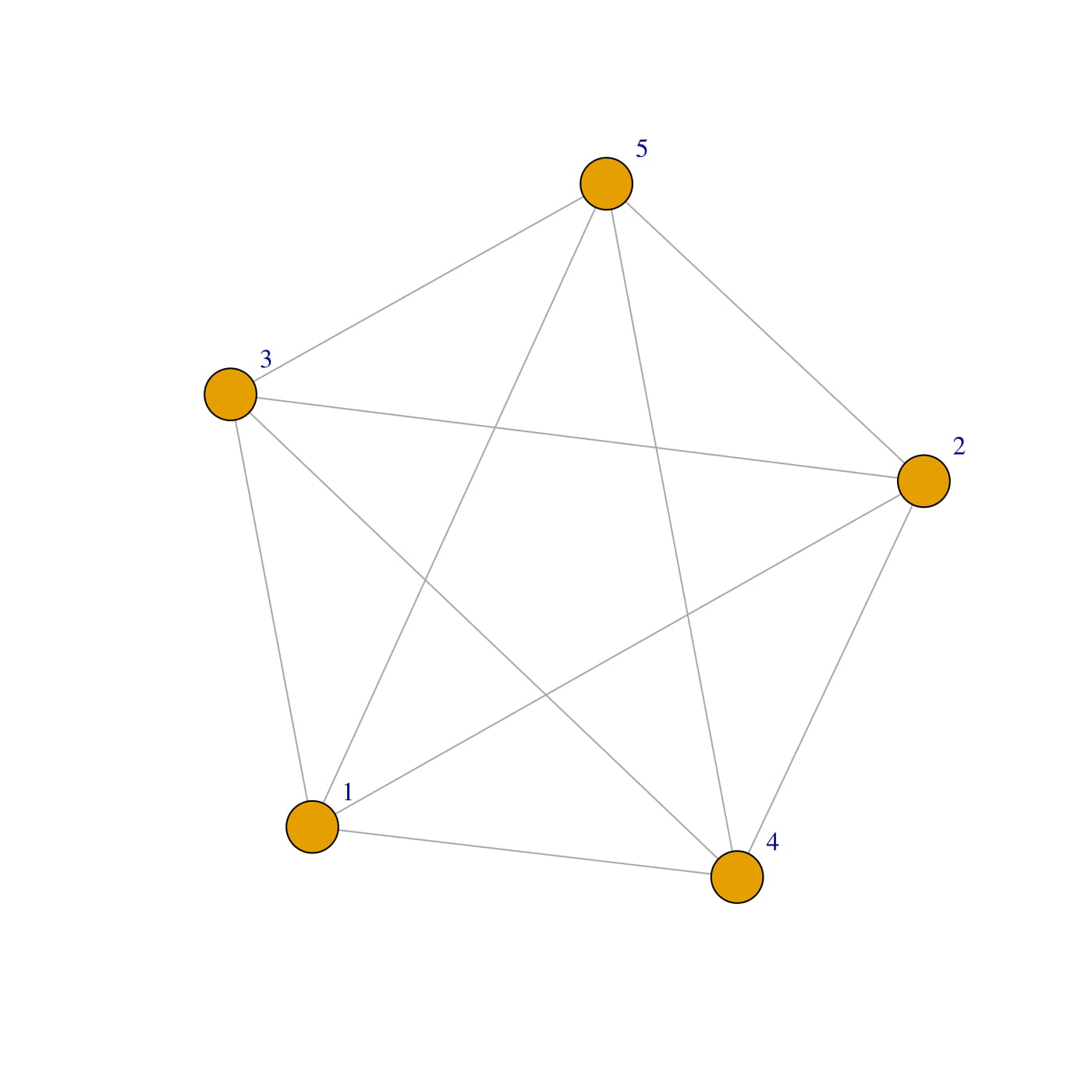}
		\caption{}
	\end{subfigure}%
	\begin{subfigure}{.33\textwidth}
		\centering
		\includegraphics[width=1.1\linewidth , height=1\linewidth]{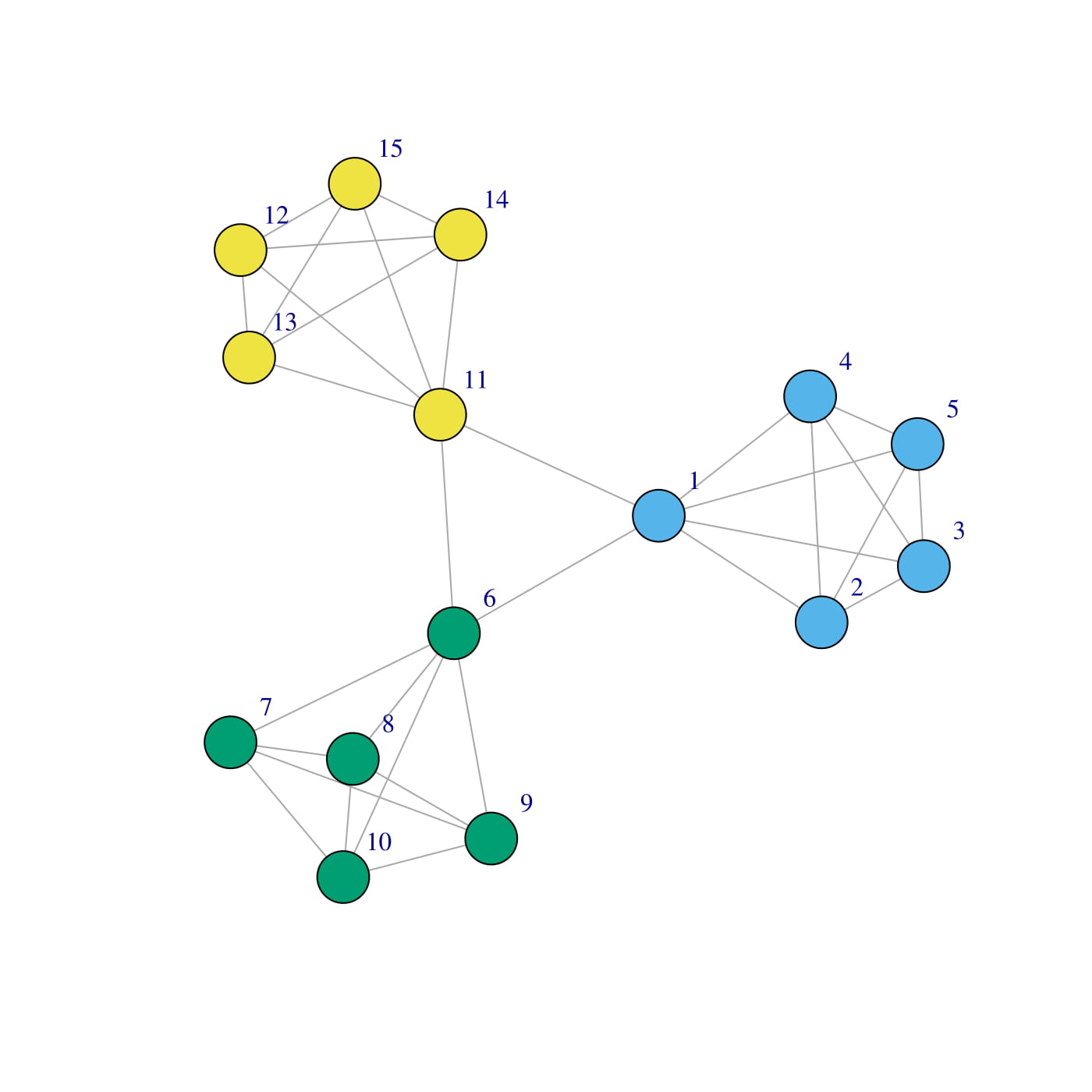}
		\caption{}
	\end{subfigure}\\
	\begin{subfigure}{.33\textwidth}
		\centering
		\includegraphics[width=1\linewidth , height=1\linewidth]{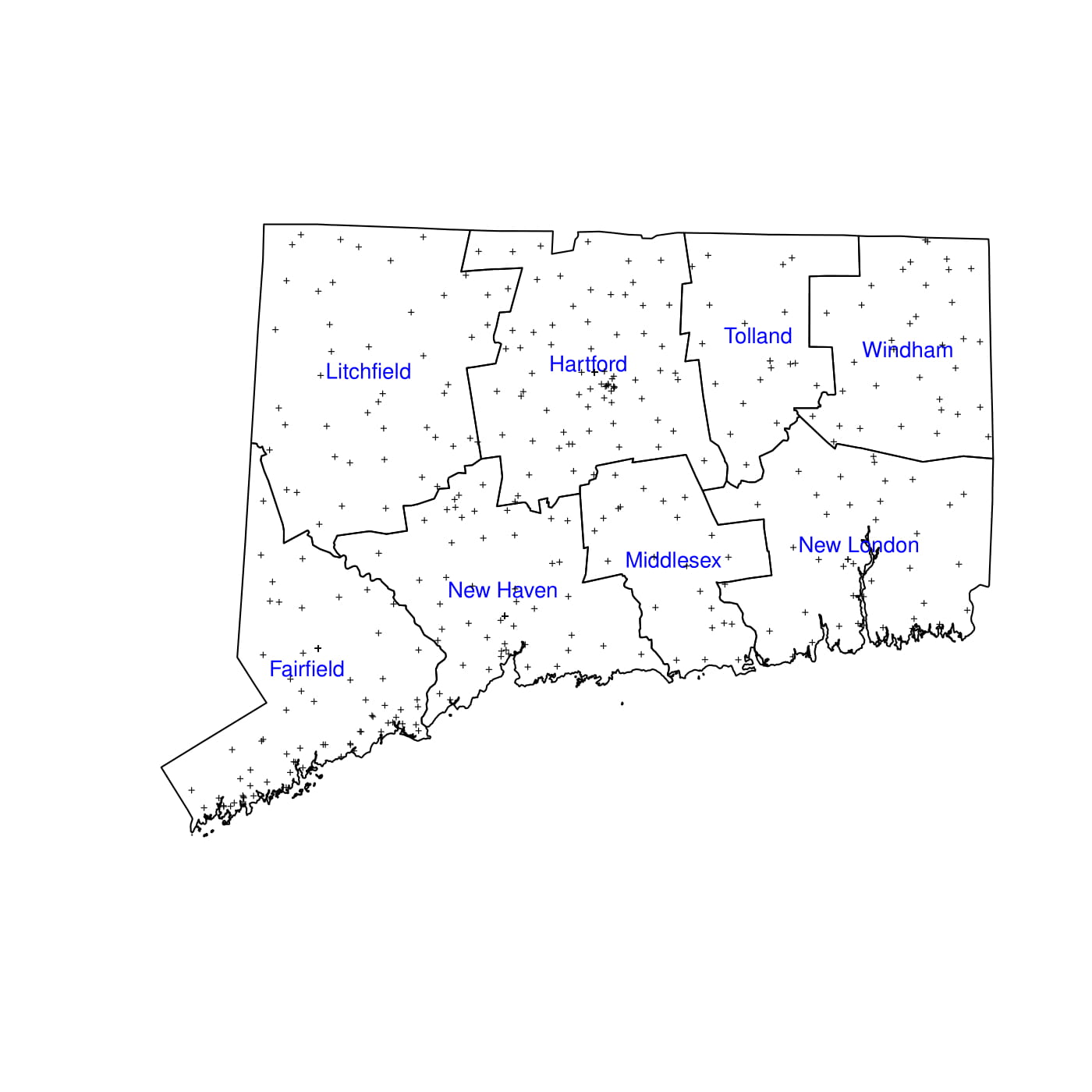}
		\caption{}
	\end{subfigure}%a
	\begin{subfigure}{.33\textwidth}
		\centering
		\includegraphics[width=1\linewidth , height=1\linewidth]{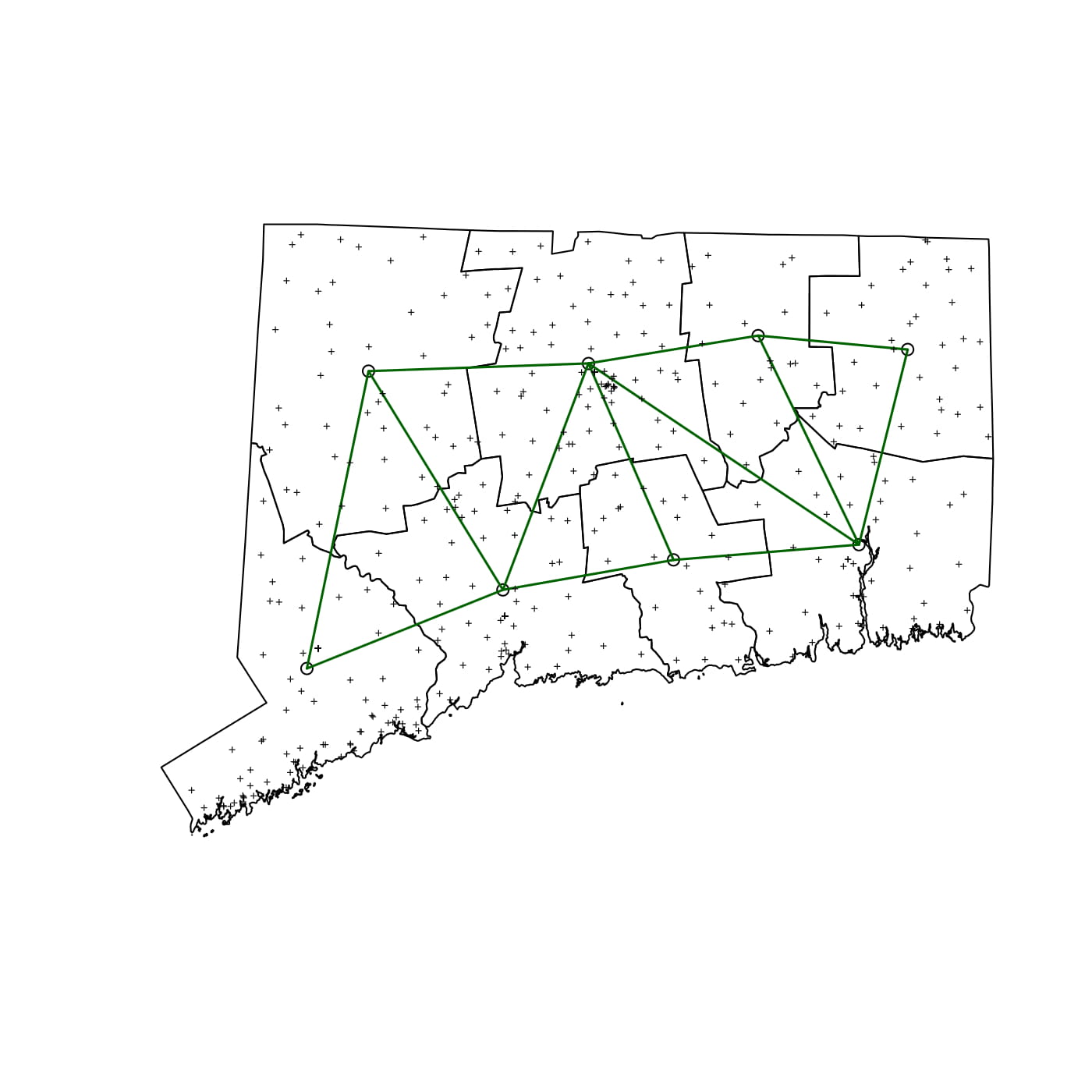}
		\caption{}
	\end{subfigure}%
	\begin{subfigure}{.33\textwidth}
		\centering
		\includegraphics[width=1.1\linewidth , height=1\linewidth]{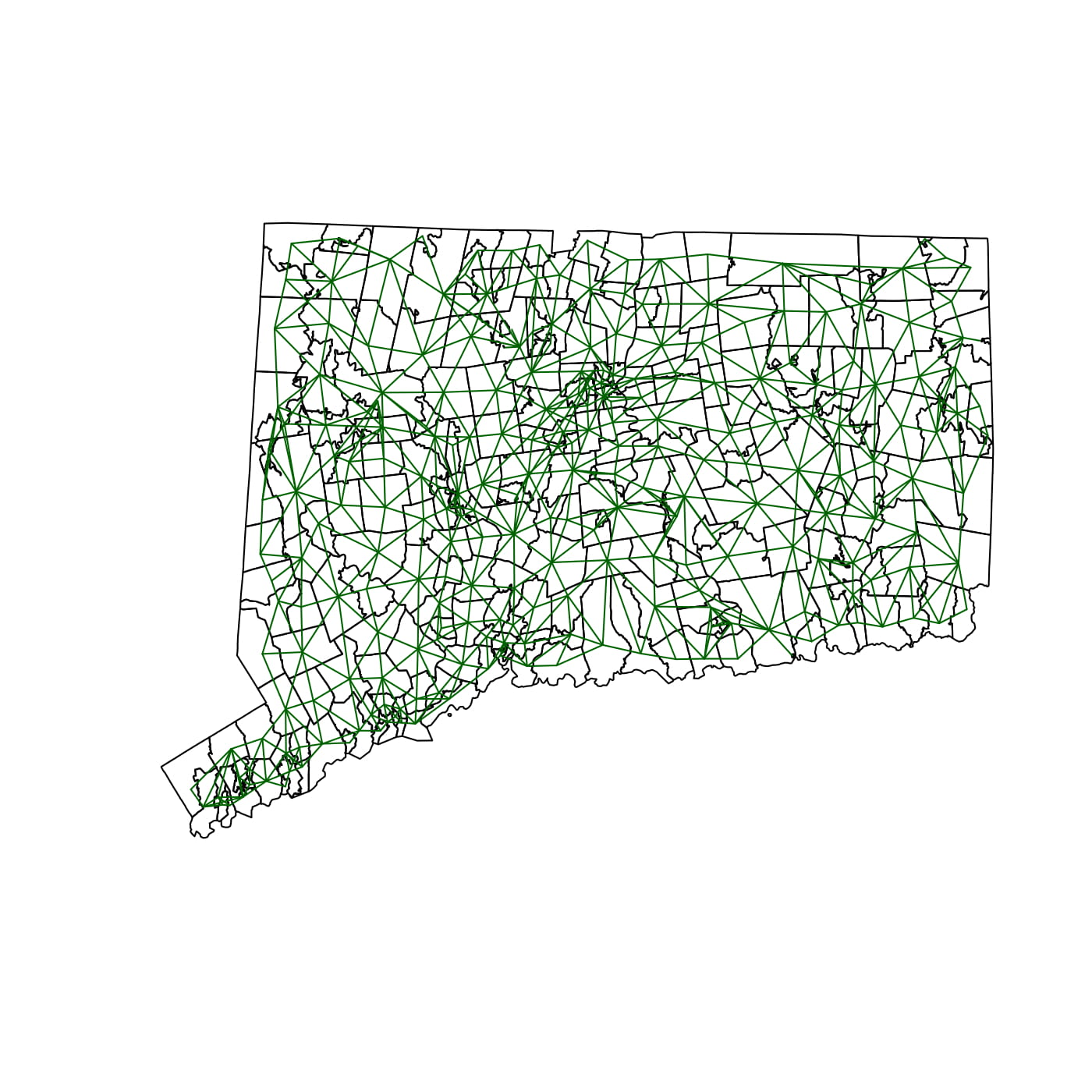}
		\caption{}
	\end{subfigure}
	\caption{Figures showing (a) a graph with 5 vertices (b) a {\em fully connected} graph with 5 vertices (c) a graph with 15 vertices with 3 fully connected sub-graphs, each with 5 vertices. Spatial plots showing the (d) 8 counties with zip codes, (e) construction of adjacency based on 8 counties, (f)  construction of adjacency based on 282 zip-codes for state of Connecticut.}
	\label{fig::graph}
\end{figure}

Spatial models have previously not been studied under a DGLM framework. In this paper we focus on introducing a spatial effect for the mean, while the DGLM allows for a varying dispersion. Specification of a spatial effect involves specifying a neighborhood structure through a undirected graph. A graph $G$ is defined as a pair $(V,E)$, consisting of vertices $V$ and edges $E$ between the vertices. If two vertices share an edge we refer to them as neighbors. The edges contain no information regarding direction producing an {\em undirected graph}. With respect to the model a graph $G$ specifies interactions between specified vertices $V$ included in the model. The first row of figure (\ref{fig::graph}) shows examples of undirected graphs. In particular (\ref{fig::graph}b) features a {fully connected graph}, i.e. starting from any vertex one can traverse the graph to reach any other vertex. Figure (\ref{fig::graph}c) shows a complex graph with 3 fully connected components which are again connected with each other. For every such graph we can create two matrices that quantify and express information in these figures namely, an adjacency and a degree matrix. If $|\cdot|$ denotes set cardinality, then adjacency and degree matrices are of order $|V|\times |V|$. For the adjacency matrix each row is specific to a vertex containing zeros for non-neighbors and ones for neighbors, while the degree matrix is a diagonal matrix containing the number of neighbors for respective vertices in its diagonal entries. Consequently, the adjacency matrix contains neighborhood information, while the degree matrix contains information regarding sparsity of the graph. Such matrices have seen widespread use to specify spatial interactions, with each vertex being a areal co-ordinate or a point reference (see \cite{cressie1992statistics}, \cite{besag1974spatial}). Edges in such scenarios are defined if two areal co-ordinate shares a boundaries. While working with zip-codes we consider the centroid for each zip code and term two zip codes as neighbors if associated polygons share boundaries. The second row of figure (\ref{fig::graph}) shows this procedure pictorially. It is evident that based on the choice of areal referencing the associated graph can prove to be fairly complicated.

The proposed approach in this paper estimates said spatial effects by adopting a penalized estimation approach. In particular we penalize the norm induced by the graph Laplacian $\mathfrak{W}=\mathfrak{W}(G)=\mathcal{D}-\mathcal{W}$, which is defined as the difference of the degree matrix $\mathcal{D}$ and adjacency matrix $\mathcal{W}$. We emphasize on joint modeling dependence of mean on covariates and the spatial domain of reference and dispersion on chosen covariates. We then proceed to develop a co-ordinate descent algorithm that models all mentioned model parameters (including the index parameter). Once such a procedure concludes, inference on estimated spatial effects proceeds via {\em krigging}, i.e. spatial interpolation (see \cite{cressie1992statistics}, \cite{stein2012interpolation}). The primary {\em assumption} behind this is that the latent unobserved surface, over $\mathbb{R}^2$, being estimated is smooth and, observed at discrete areal locations or co-ordinates. This also facilitates prediction at chosen unobserved co-ordinates (see \cite{rmbacite}). Including spatial effects within any modeling framework introduces over-dispersion into the model accounting for variability offered by the underlying spatial domain of reference. Naturally not specifying this translates to bias in estimated model coefficients for mean and dispersion models.

In section (\ref{sec::tweedie}) we formally introduce the Tweedie ED family and DGLMs in that context. We then proceed to fit proposed spatial DGLM by formulating it as an optimization problem, section (\ref{sec::opt-prob}) discusses details regarding its formulation. This includes a proposed algorithm that can be used for fitting these models for members in the family. We refrain from introducing unnecessary restrictions on the proposed model framework, which results in a statement of minimal necessary criteria required on probability densities and link functions for said algorithm to be applicable. Details and challenges for individual members of the Tweedie family are also considered, demonstrating that suggested approach is applicable to all members of the family. Section (\ref{sec::sim}) considers evaluating performance of suggested approach in comparison to state-of-the-art penalties like the ridge penalty to showcase efficacy of proposed approach. Section (\ref{sec::real-apply}) considers fitting a spatial DGLM to automobile collision insurance losses using proposed approach for the state of Connecticut, in the year 2008. Finally, section (\ref{sec::remarks}) concludes the paper and discusses possibilities for future research in hindsight of proposed developments.

\section{Tweedie ED family and DGLMs}\label{sec::tweedie}

We begin with some notations, let $y_{ij}$, $\mu_{ij}$, $\phi_{ij}$ be the response, mean and dispersion respectively and $x_{ij}\trans$, $r_{ij}\trans$ and $z_{ij}\trans$ denote observed covariate vectors for $j$-th observation at the $i$-th vertex, where $j=1,\ldots,n_i$ and $i=1,\ldots, L$, with $\sum_{i=1}^{L}n_i=N$ being the total number of observations. To make notations less cumbersome, when convenient, subscripts are omitted in following discussion.

Tweedie densities belong to the more general ED family whose probability density function is given by,
\begin{align}\label{eq::edm-dens}
f(y;\theta,\phi) = a(y,\phi) \exp\Big\{\phi^{-1}\big(y\theta-\kappa(\theta)\big)\Big\}.
\end{align}
where $\theta$, $\phi$ are the natural or, canonical and dispersion parameter respectively and $\kappa(\theta)$ the cumulant function.  The Tweedie family specifies an additional index parameter $p$, distinguishing between included members from the ED family. If we denote the mean by $\mu=E(y)$ and variance by, $\text{Var}(y)$, then $\mu=\kappa'(\theta)$ and $\text{Var}(y)=\phi\kappa''(\theta)$. Using $\mu$, $\phi$ and $p$ we write $y\sim Tw(\mu,\phi,p)$. Probability density function in eq. (\ref{eq::edm-dens}) can also be expressed using unit-deviance $d(y,\mu)=-2\int_{y}^{\mu}\frac{y-u}{V(u)}\mathrm{d}u$; table (\ref{tab::tweedie-pars}) lists out closed forms for some commonly used members in the Tweedie ED family. A special feature of the ED family is a relationship between mean and variance, that is specified through a variance function $V(\mu)$ unique to each member. For members of the Tweedie ED family, $\text{Var}(y)=\phi\kappa''(\theta)=\phi V(\mu)=\phi\mu^p$. The value of $p$ and therefore expression $V(\mu)$ uniquely determines the member in question. 

DGLMs are a generalization of GLMs that provide a more general framework by allowing the dispersion $\phi$ to vary across observations. For appropriate choice of link functions $g_1$ and $g_2$ a DGLM is formulated as
\begin{align}\label{eq::DGLM}
g_1(\mu) = x\trans\bbeta &,& \mathrm{Var}(y) = \phi V(\mu) &,& g_2(\phi) = z\trans\bgamma.
\end{align}
Additionally, a graphical DGLM starts by specifying additional unobserved parameters $\balpha(G)$, with associated dependence structure being explained by graph $G$. Henceforth, we shall refer to $\balpha(G)$ as $\balpha$. This specification extends model in eq. (\ref{eq::DGLM}) to
\begin{align}\label{eq::graph-DGLM}
g_1(\mu) = x\trans\bbeta+r\trans\balpha &,& \mathrm{Var}(y) = \phi V(\mu) &,& g_2(\phi) = z\trans\bgamma.
\end{align}
We denote $x\trans$, $z\trans$ as covariate vectors corresponding to mean and dispersion respectively while, $r\trans$ is a $L\times 1$ vector consisting of zeros, except for 1 in one entry indicating a particular edge in the graph. Another advantage of specifying $\balpha$ is observed through joint modeling of $\bbeta$ and $\balpha$. Specification of the dependence structure using the graph for $\balpha$ affects the quality of estimation for $\bbeta$ through bias adjustment within the mean model. Denoting the inverse maps $g_i^{-1}=h_i$ for $i=1,2$, $(\kappa')^{-1}=k$ and the function composition $k_{h_1}=k\circ h_1=k(h_1(\cdot),p)$, the negative log-likelihood for the family can be expressed as
\begin{equation}
	\begin{aligned}\label{eq::neg-log-lik-tw-ed}
		\ell(\bbeta,\balpha,\bgamma;y,p)=-\sum\limits_{i=1}^{L}\sum\limits_{j=1}^{n_i}\mfrac{1}{h_2(z_{ij}\trans\bgamma)}\Big[y_{ij}k_{h_1}(x_{ij}\trans\bbeta+r_{ij}\trans\balpha)-\kappa\big(k_{h_1}(x_{ij}\trans\bbeta+r_{ij}\trans\balpha)\big)\Big]\\+\log a(y_{ij},h_2(z_{ij}\trans\bgamma),p).
	\end{aligned}
\end{equation}

\begin{table}[t]
	\renewcommand{\arraystretch}{1.8}
	\centering
	\caption{Commonly used members of the Tweedie family of distributions with their index parameter ($p$), variance function ($V(\mu)$), cumulant function ($\kappa(\theta)$), canonical parameter ($\theta$), dispersion ($\phi$), deviance ($d(y,\mu)$), normalizing constant ($a(y,\phi)$), support ($S$), and respective parameter spaces for mean ($\mathit{\Omega}$) and the natural parameter ($\Theta$).}\label{tab::tweedie-pars}
	\resizebox{\linewidth}{!}{
		\begin{tabular}{|l|cccccccccc|}
			\hline
			\multirow{2}{*}{Tweedie EDMs} & \multirow{2}{*}{$p$} & \multirow{2}{*}{$V(\mu)$} & \multirow{2}{*}{$\kappa(\theta)$} & \multirow{2}{*}{$\theta$} & \multirow{2}{*}{$\phi$} & \multirow{2}{*}{$d(y,\mu)$} & \multirow{2}{*}{$a(y,\phi)$} & \multirow{2}{*}{$S$} & \multirow{2}{*}{$\mathit{\Omega}$} & \multirow{2}{*}{$\Theta$} \\
			&&&&&&&&&&\\
			\hline
			Normal & 0 & 1 & $\mfrac{\theta^2}{2}$ & $\mu$ & $\sigma^2$ & $(y-\mu)^2$ & $\sqrt{2\pi\phi}$ & $\mathbb{R}$ & $\mathbb{R}$ & $\mathbb{R}$\\
			Poisson & 1 & $\mu$ & $\exp(\theta)$ & $\log(\mu)$ & 1 & $2\Big\{y\log\mfrac{y}{\mu}-(y-\mu)\Big\}$ & $\mfrac{1}{y!}$ & $\mathbb{N}\cup\{0\}$ & $\mathbb{R}^{+}$ & $\mathbb{R}$\\
			Poisson-gamma & $(1,2)$ & $\mu^p$ & $\mfrac{\{(1-p)\theta\}^{(2-p)/(1-p)}}{2-p}$ & $\mfrac{\mu^{1-p}}{1-p}$ & $\phi$ & $2\Big\{\mfrac{\max(y,0)^{2-p}}{(1-p)(2-p)}-\mfrac{y\mu^{1-p}}{1-p}+\mfrac{\mu^{2-p}}{2-p}\Big\}$ & -- & $\mathbb{R}^{+}\cup\{0\}$ & $\mathbb{R}^{+}$ & $\mathbb{R}^{-}$\\
			Gamma & 2 & $\mu^2$ & $-\log(-\theta)$ & $-\mfrac{1}{\mu}$ & $\phi$ & $2\Big\{-\log\mfrac{y}{\mu}+\mfrac{y-\mu}{\mu}\Big\}$ & $\mfrac{\phi^{-1/\phi} y^{1/\phi-1}}{\Gamma(1/\phi)}$ & $\mathbb{R}^{+}$ & $\mathbb{R}^{+}$ & $\mathbb{R}$\\
			Inverse Gaussian & 3 & $\mu^3$ & $-\sqrt{-2\theta}$ & $-\mfrac{1}{2\mu^2}$ & $\phi$ & $\mfrac{(y-\mu)^2}{\mu^2y}$ & $(2\pi y^3\phi)^{-1/2}$ & $\mathbb{R}^{+}$ & $\mathbb{R}^{+}$ & $\mathbb{R}$\\
			\hline
		\end{tabular}
	}
\end{table}

As is evidenced from table (\ref{tab::tweedie-pars}) depending on $p$, dispersion can be constant, for example Poisson models; in that case we choose $g_2(\cdot)=1$ resulting in a simple GLM for $\mu$. Furthermore, depending on $p$ the density may lack a closed-form expression for associated normalizing factor, resulting in complex model structure for instance CP-g models. While dealing with CP-g densities we operate under $1<p<2$. Explicitly, CP-g densities have the following form
\begin{subequations}
	\begin{align}\label{eq::cp-g-dens-1}
	f(y;\theta,\phi,p) = a(y,\phi,p)\exp\Bigg[\phi^{-1}\Bigg\{y\theta-\mfrac{\big((1-p)\theta\big)^{-\xi}}{2-p}\Bigg\}\Bigg] &,& \theta=\mfrac{\mu^{1-p}}{1-p} &,& 1<p<2,
	\end{align}
	where we have,
	\begin{align}\label{eq::cp-g-dens-2}
	\xi=\mfrac{2-p}{p-1}&,& a(y,\phi,p)=\begin{cases}
	y^{-1}\sum\limits_{k=1}^{\infty}a_k(y,p)\phi^{-k(1+\xi)} & y>0\\
	1&y=0
	\end{cases} &,& \scalemath{0.9}{a_k(y,p)=\mfrac{y^{k\xi}(p-1)^{-k\xi}}{(2-p)^{k}k!\Gamma(k\xi)}}.
	\end{align}
\end{subequations}
Expression in eq. (\ref{eq::cp-g-dens-2}) is obtained by marginalizing out the associated random Poisson count from their joint density followed by equating resulting expression with eq. (\ref{eq::edm-dens}). This results in an infinite sum representation for $a(y,\phi,p)$, which was identified as an example of the generalized Bessel function, lacking a closed form. This translates to a lack of normalizing factor for the density, and the need for an approximation to the infinite sum (in eq. \ref{eq::cp-g-dens-2}) to make the density computationally viable. A fairly accurate approximation for the infinite sum is obtained through locating an approximate mode for terms $a_k(y,p)\phi^{-k(1+\xi)}$ at $k_{\max}=y^{2-p}/(2-p)\phi$, followed by selecting an appropriate radius of inclusion for terms around $k_{\max}$, for {\em every} $y$ (for details see section 4, \cite{dunn2005series}). Modified {\em saddle-point approximations} provide an analytically simpler approach (see \cite{dunn2018generalized}) by approximating densities for the ED family using,
\begin{align}\label{eq::cp-g-dens-saddle}
\tilde{f}(y;\theta,\phi,p) = b(y,\phi,p)\exp\Big[-\mfrac{d(y,\mu)}{2\phi}\Big] &,& b(y,\phi,p)\approx(2\pi\phi V(y))^{-1/2},
\end{align}
 for Tweedie EDs $V(y)=y^pI(y>0)+(y+\epsilon_0)^pI(y=0)$. For instance, using eqs. (\ref{eq::cp-g-dens-1}), (\ref{eq::cp-g-dens-2}) and (\ref{eq::graph-DGLM}) with $g_1=g_2=\log(\cdot)$, we have the following {\em negative log-likelihood},
\begin{subequations}
	\begin{gather}
	C_{ij}=y_{ij}^{-1}\sum\limits_{k=1}^{\infty}a_k(y_{ij},p)e^{-k(1+\xi)z_{ij}\trans\small \bgamma}\cdot I(y_{ij}>0)+1\cdot I(y_{ij}=0),\label{eq::cpg-neg-log-lik-bessel-i}\\
	\ell(\bbeta,\balpha,\bgamma;p,y) = -\sum\limits_{i}\sum\limits_{j}e^{-z\trans_{ij}\small \bgamma}\Bigg[y_{ij}\mfrac{e^{(1-p)(x_{ij}\trans{\small \bbeta}+r_{ij}\trans{\small \balpha})}}{1-p}-\mfrac{e^{(2-p)(x_{ij}\trans{\small \bbeta}+r_{ij}\trans{\small \balpha})}}{2-p}\Bigg]+\log C_{ij},\label{eq::cpg-neg-log-lik-bessel-ii}
	\end{gather}
	equivalently under modified saddle-point approximation,
	\begin{gather}
	C_{ij}=(2\pi y_{ij}^p e^{-z_{ij}\trans\small\bgamma})\cdot I(y_{ij}>0)+(2\pi \epsilon_0^p e^{-z_{ij}\trans\small\bgamma})\cdot I(y_{ij}=0),\label{eq::cpg-neg-log-lik-saddle-i}\\
	\scalemath{0.9}{\ell(\bbeta,\balpha,\bgamma;p,y) = -\sum\limits_{i}\sum\limits_{j}e^{-z\trans_{ij}\small \bgamma}\Bigg[y_{ij}\mfrac{y_{ij}^{1-p}-e^{(1-p)(x_{ij}\trans{\small \bbeta}+r_{ij}\trans{\small \balpha})}}{1-p}-\mfrac{y_{ij}^{2-p}-e^{(2-p)(x_{ij}\trans{\small \bbeta}+r_{ij}\trans{\small \balpha})}}{2-p}\Bigg]+\log C_{ij}},\label{eq::cpg-neg-log-lik-saddle-ii}
	\end{gather} 
\end{subequations}
where $I(\cdot)$ is an indicator function. Naturally, choice of $C_{ij}$ can be made based on chosen method of approximation. Evidently, $\epsilon_0$ is chosen to be small for saddle point approximations.

\section{Optimization Problem}\label{sec::opt-prob}

Denoting $\Theta=(\bbeta,\balpha,\bgamma)\trans$, model in eqn. (\ref{eq::graph-DGLM}) is fitted by solving the optimization problem,
\begin{subequations}
	\begin{align}\label{eq::opt-prob-i}
	\left\{\widehat{\Theta}, \widehat{p}\right\} = \mathrm{arg}\min\limits_{\Theta,p} F\left(\Theta,p\right) &,& F(\Theta,p) = \ell(\Theta,p) + P ({\bf A}\Theta; \lambda_1, \lambda_2).
	\end{align}
	Here $p$ is an additional index parameter associated with model selection. The penalty function is
	\begin{align}\label{eq::opt-prob-ii}
		P ({\bf A}\Theta; \lambda_1, \lambda_2) = \mfrac{1}{2}({\bf A}\Theta)\trans\big[\lambda_1{\bf I}_0+\lambda_2{\bf W}_0\big]{\bf A}\Theta,
	\end{align}
	where ${\bf A}$ is a vector having the same dimension as $\Theta$.%, with possible linear constraints,
%	\begin{align}\label{eq::lin-constr}
%		{\bf B}\trans\Theta={\bf d} &,& \arraycolsep=4.0pt\def\arraystretch{1.5} {\bf B}=\left[\begin{array}{c:c} 
%		{\bf B}_\eta & O\\\hdashline
%		O&{\bf B}_\gamma
%		\end{array}\right]&,& r({\bf B})=r({\bf B}_\eta)+r({\bf B}_\gamma)=q,
%	\end{align}
\end{subequations}
 %with $q<(\dim(\bbeta)+L)+\dim(\bgamma)$. 
 Penalty function $P ({\bf A}\Theta; \lambda_1, \lambda_2)$ in its most general form, consists of two parts (a) first part consisting of ${\bf I}_0$, denotes a ridge penalization on $\Theta$, (b) second part consists of ${\bf W}_0$, which penalizes a graph derivative, the Laplacian $\mathfrak{W}$ in this case,  promoting clustering on the graph. Note that if required, entries of ${\bf A}$ can be made zero to {\em not} penalize a particular component of $\Theta$. For instance, if we choose to only penalize $\balpha$ we have,
\begin{align}\label{eq::config-1}
	{\bf A}=(0,\ldots,0,\underbrace{1,\ldots,1}_{L},0\ldots,0)&,& {\bf I}_0={\bf I}_L &,& {\bf W}_0=\mathfrak{W},
\end{align}
additionally, choosing to have a ridge penalty on $\bbeta$ and $\bgamma$ results in,
\begin{align}\label{eq::config-2}
{\bf A}=(\overbrace{1,\ldots,1}^{\dim(\small \bbeta)},\underbrace{1,\ldots,1}_L,\overbrace{1\ldots,1}^{\dim(\small \bgamma)})&,& {\bf I}_0={\bf I}_{\dim(\small\bbeta)+L+\dim(\small\bgamma)} &,& \arraycolsep=5.0pt\def\arraystretch{1.5} {\bf W}_0=\left[ \begin{array} {c:c:c} O &  O &  O \\\hdashline   O & \mathfrak{W} & O\\\hdashline O & O & O\end{array}\right],
\end{align}
where ${\bf W}_0$ is partitioned appropriately. Furthermore, depending on the nature of covariates, ${\bf W_0}$ need not be as sparse as in eq. (\ref{eq::config-2}). Interdependence between covariates for mean and dispersion models can be penalized in the diagonal blocks. While off-diagonal blocks can be used to allow for penalizing dependence of groups of covariates on the graph, dependence of covariates across the mean and dispersion models can also be accommodated. We shall solve the optimization problem using configurations in eq. (\ref{eq::config-1}) followed by discussing necessary modifications required to extend formulated solution for the configuration in eq. (\ref{eq::config-2}). 

 Keeping $p$ fixed, we partition $\Theta=(\bbeta,\balpha, \bgamma)\trans=(\bleta\trans,\bgamma)\trans$, where $\bleta=(\bbeta,\balpha)\trans$. If $\dim(\bbeta)=k_\beta$, $\dim(\bgamma)=k_\gamma$, i.e. $\bbeta$, $\balpha$ are $k_\beta\times 1$ possibly including intercepts, let us denote $(k_\beta+L)\times 1$ and $k_\gamma\times1$ gradient vectors as $\nabla_1(\bleta|\bgamma)$ and $\nabla_1(\bgamma|\bleta)$ and associated Hessians of order $(k_\beta+L)\times (k_\beta+L)$ and $k_\gamma\times k_\gamma$ by $\nabla_2(\bleta|\bgamma)$ and $\nabla_2(\bgamma|\bleta)$ respectively. $\nabla_1(\bleta|\bgamma)$ and $\nabla_2(\bleta|\bgamma)$ are partitioned according to $\bleta$. The optimization problem in eq. (\ref{eq::opt-prob-i}, \ref{eq::opt-prob-ii}) is solved by iteratively solving three sub-problems in $\bleta$, $\bgamma$ and $p$,
 \begin{subequations}
 	\begin{gather}
	 	\scalemath{0.9}{\argmin_{\bleta^*}~\ell(\Theta,p)+(\bleta^*-\bleta)\trans\nabla_1(\bleta|\bgamma)+\mfrac{c_1}{2}(\bleta^*-\bleta)\trans \nabla_2(\bleta|\bgamma)(\bleta^*-\bleta)+P({\bf A}\Theta;\lambda_1,\lambda_2)},\label{eq::opt-mean-prob}\\%+\Delta(\Theta^*)
	 	\scalemath{0.9}{\argmin_{\bgamma^*}~\ell(\Theta,p)+(\bgamma^*-\bgamma)\trans\nabla_1(\bgamma|\bleta^*)+\mfrac{c_2}{2}(\bgamma^*-\bgamma)\trans \nabla_2(\bgamma|\bleta^*)(\bgamma^*-\bgamma)+P({\bf A}\Theta;\lambda_1,\lambda_2)},\label{eq::opt-disp-prob}\\ %+\Delta(\Theta^*)
	 	\argmin_{p}\ell(\Theta^*,p),\label{eq::opt-index-prob}
 	\end{gather}
 	where
 	%\begin{align}
	 %	\Delta(\Theta)=\Delta(\Theta; {\bf A},{\bf B},{\bf c}, {\bf d})=P({\bf A}\Theta;\lambda_1,\lambda_2)+{\bf c}\trans({\bf B}\trans\Theta-{\bf d}),
 	%\end{align}
 \end{subequations}%
 $c_1$, $c_2$ are constants.% and ${\bf c}$ is a $q$ dimensional vector of Lagrange multipliers. 
 Note that using ${\bf A}$ in eqs. (\ref{eq::config-1}) and (\ref{eq::config-2}) automatically adjusts penalization for optimization problems in eqs. (\ref{eq::opt-mean-prob}) and (\ref{eq::opt-disp-prob}) respectively. Under configuration in eq. (\ref{eq::config-1}) solution to the problems are,
 \begin{subequations}
 	\begin{gather}
 	\bleta^{*}= [\lambda_1{\bf I_0}+\lambda_2{\bf W_0}+c_1\nabla_2(\bleta|\bgamma)]^{-1}\{c_1\nabla_2(\bleta|\bgamma)\bleta-\nabla_1(\bleta|\bgamma)\},\label{eq::sol-eta}\\%-{\bf B}_\eta{\bf c}_\eta
 	\bgamma^{*}= \bgamma-\mfrac{1}{c_2}\nabla_2^{-1}(\bgamma|\bleta^*)\{\nabla_1(\bgamma|\bleta^*)\}\label{eq::sol-gamma},%+{\bf B}_\gamma {\bf c}_\gamma
 	\end{gather}
 \end{subequations}
 with 
 \begin{align*}
 {\bf I_0} = \arraycolsep=4.0pt\def\arraystretch{1.5}\left[ \begin{array} {c:c} O & O \\\hdashline   O & I_L\end{array}\right]&,& {\bf W_0} = \arraycolsep=4.0pt\def\arraystretch{1.5}\left[ \begin{array} {c:c}O & O \\\hdashline   O & \mathfrak{W}\end{array}\right],
 \end{align*}
 are partitioned matrices of order $(k_\beta+L)\times (k_\beta+L)$. %Solving ${\bf B}_\eta\trans\bleta^*={\bf d}_\eta$, using solution from eq. (\ref{eq::sol-eta}) while denoting ${\bf S}= \lambda_1{\bf I_0}+\lambda_2{\bf W_0}+c_1\nabla_2(\bleta|\bgamma)$ and $\bleta_0 = {\bf S}^{-1}\{c_1\nabla_2(\bleta|\bgamma)\bleta-\nabla_1(\bleta|\bgamma)\}$,
 %\begin{align*}
 %{\bf c}_\eta = \big({\bf B}_\eta\trans{\bf S}^{-1}{\bf B}_\eta\big)^{-1}\big({\bf B}_\eta\trans\bleta_0-{\bf d}_\eta\big).
 %\end{align*} 
 %Under ${\bf B}_\eta\trans\bleta={\bf d}_\eta$
  There exists $c_1$ such that,
 \begin{eqnarray}
 F(\bleta^*|\bgamma,p)-F(\bleta|\bgamma,p)&=&\ell(\bleta^*|\bgamma,p) + P (\bleta^*; \lambda_1, \lambda_2)-\ell(\bleta|\bgamma,p)-P (\bleta; \lambda_1, \lambda_2),\nonumber\\
 &\approx& \ell(\bleta|\bgamma,p)+(\bleta^*-\bleta)\trans\nabla_1(\bleta|\bgamma)+\mfrac{1}{2}(\bleta^*-\bleta)\trans c_1\nabla_2(\bleta|\bgamma)(\bleta^*-\bleta)\nonumber\\
 &&+P(\bleta^*;\lambda_1,\lambda_2)-\ell(\bleta|\bgamma,p)-P(\bleta;\lambda_1,\lambda_2) \leq 0.\nonumber
 \end{eqnarray}
 The last inequality is proved in Appendix A. %In eq. (\ref{eq::sol-gamma}) solution for ${\bf c}_\gamma$ is obtained by solving ${\bf B}_\gamma\trans\bgamma^*={\bf d}_\gamma$. Furthermore under ${\bf B}_\gamma\trans\bgamma={\bf d}_\gamma$, 
 There exists constant $c_2$ such that,
 \begin{align}
 F(\bgamma^*|\bleta^*,p)-F(\bgamma|\bleta^*,p)&=\ell(\bgamma^*|\bleta^*,p)-\ell(\bgamma|\bleta^*,p),\nonumber\\
 ~~\approx \ell(\bgamma|\bleta^*,p)&+(\bgamma^*-\bgamma)\trans\nabla_1(\bgamma|\bleta^*)+\mfrac{1}{2}(\bgamma^*-\bgamma)\trans c_2\nabla_2(\bgamma|\bleta^*)(\bgamma^*-\bgamma)
 -\ell(\bgamma|\bleta^*,p),\nonumber\\
 &=-\mfrac{1}{2c_2}\nabla\trans_1(\bgamma|\bleta^*)\nabla^{-1}_2(\bgamma|\bleta^*)\nabla_1(\bgamma|\bleta^*)\leq 0.\label{eq::disp-cond}
 \end{align}
 The second equality is obtained by substituting solution in eq. (\ref{eq::sol-gamma}). Positive or negative definiteness of $\nabla^{-1}_2(\bgamma|\bleta^*)$ in eq. (\ref{eq::disp-cond}), facilitates choice of $c_2$. Under ridge penalization for configuration in eq. (\ref{eq::config-2}), the solution for eqs. (\ref{eq::opt-mean-prob}) and (\ref{eq::opt-disp-prob}) are,
 \begin{subequations}
 	\begin{gather}
 	\bleta^{*}= [\lambda_1{\bf I}_{k_\beta+L}+\lambda_2{\bf W_0}+c_1\nabla_2(\bleta|\bgamma)]^{-1}\{c_1\nabla_2(\bleta|\bgamma)\bleta-\nabla_1(\bleta|\bgamma)\},\label{eq::sol-ridge-eta}\\%-{\bf B}_\eta{\bf c}_\eta
 	\bgamma^{*}= \big(\lambda_1{\bf I}_{k_\gamma}+c_2\nabla_2(\bgamma|\bleta^*)\big)^{-1}\big(c_2\nabla_2(\bgamma|\bleta^*)\bgamma-\nabla_1(\bgamma|\bleta^*)\big)\label{eq::sol-ridge-gamma},%-{\bf B}_\gamma{\bf c}_\gamma
 	\end{gather}
 \end{subequations}
 Following similar logic for proof shown in Appendix A, %solving for ${\bf c}_\eta$, ${\bf c}_\gamma$ 
 and choosing $c_1$, $c_2$ such that $c_1\nabla_2(\bleta|\bgamma)$ and $c_2\nabla_2(\bgamma|\bleta^*)$ are positive semi-definite (p.s.d) respectively, ensures $F(\Theta,p)-F(\Theta^*,p)\geq 0$. Finally we update $p$ by maximizing the profile likelihood. Proceedings are concisely summarized in algorithm (\ref{algo::md-1}), accompanied by theorem (\ref{th::conv}) (proof shown in Appendix A) proving convergence for proposed solutions.
 Theorem (\ref{th::conv}) shows that under proper choice of scaling constants $c_1$ and $c_2$, the objective function is guaranteed to decrease for all $\lambda_1> 0$. The choice of a convergence criteria for the algorithm is based on the objective function $F(\Theta,p)$, equivalently this implies $||\Theta-\Theta^*||^2_2\leq 2\epsilon/\lambda_1$, i.e., for arbitrarily small $\epsilon$ the difference in Euclidean norm for estimates can be made arbitrarily small.\\
 
 \begin{algorithm}[H]
 	\SetAlgoLined
 	Initialize coefficients $\Theta=(\bbeta,\balpha,\bgamma)\trans=(\bleta,\bgamma)\trans$, index parameter $p$ and set Laplacian $\mathfrak{W}$ for graph\;
 	Compute initial $F(\Theta,p)$ and set $\epsilon_0$;
 	
 	\Repeat{Convergence of $F(\Theta,p)$, i.e. $\epsilon^*<\epsilon_0$}{
 		Compute $\nabla_1(\bleta|\bgamma)$ and $\nabla_2(\bleta|\bgamma)$ from eqs. (\ref{eq::mean-d1}) and (\ref{eq::mean-d2})\;
 		-- compute scaling $c_1$\;
 		\eIf{ridge penalty on $\bbeta$}{
 			Update $\bleta^*$ using eq. (\ref{eq::sol-ridge-eta})\;
 		}{
 			Update $\bleta^*$ using eq. (\ref{eq::sol-eta})\;
 		}
 		Update $D_{ij}(\bleta) \gets D_{ij}(\bleta^*)$	in eq. (\ref{eq::tw-dev})\;
 		Compute $\nabla_1(\bgamma|\bleta^*)$ and $\nabla_2(\bgamma|\bleta^*)$ from eqs. (\ref{eq::disp-d1}) and (\ref{eq::disp-d2})\;
 		-- compute scaling $c_2$\;
 		\eIf{ridge penalty on $\bgamma$}{
 			Update $\bgamma^*$ using eq. (\ref{eq::sol-ridge-gamma})\;
 		}{
 			Update $\bgamma^*$ using eq. (\ref{eq::sol-gamma})\;
 		}
 		Update $p^* \gets \argmin_{p}\ell(\Theta^*,p)$\;
 		Compute $\epsilon^*=F(\Theta,p)-F(\Theta^*,p^*)$\;
 		Set $\Theta \gets \Theta^*=(\bleta^*,\bgamma^*)\trans$, $p \gets p^*$\;
 	}
 	Return $\Theta$, $p$.
 	\caption{\small Algorithm for fitting a graphical CP-g DGLM through penalization of the graph Laplacian, accompanied by a ridge penalty for respective model effects.}\label{algo::md-1}
\end{algorithm}

\begin{mythe}\label{th::conv}
	For appropriately chosen constants $c_1$ and $c_2$, under configuration in eq. (\ref{eq::config-1}) and $\lambda_1>0$, sequence $\Theta^*=\{(\bleta^*,\bgamma^*)\trans\}$ from eqs. (\ref{eq::sol-eta}) and (\ref{eq::sol-gamma}) satisfy
	\begin{align*}
	F(\Theta,p)-F(\Theta^*,p^*)\geq F(\Theta,p)-F(\Theta^*,p)=F(\bleta|\bgamma,p)-F(\bleta^*|\bgamma,p)+F(\bgamma|\bleta^*,p)-F(\bgamma^*|\bleta^*,p)\\
	\geq \mfrac{\lambda_1}{2}(\bleta^*-\bleta)\trans{\bf I_0}(\bleta^*-\bleta)=\mfrac{\lambda_1}{2}||\balpha^*-\balpha||_2^2.
	\end{align*}
	Furthermore, for solutions in eqs. (\ref{eq::sol-ridge-eta}) and (\ref{eq::sol-ridge-gamma}) under configuration in eq. (\ref{eq::config-2}), we have
	\begin{align*}
	F(\Theta,p)-F(\Theta^*,p^*)\geq F(\Theta,p)-F(\Theta^*,p)\geq \mfrac{\lambda_1}{2}||\Theta^*-\Theta||_2^2,
	\end{align*}
	where $||\cdot||_2$ is the Euclidean norm.
\end{mythe}

%To justify practical need for having linear constraints, let us consider the problem in eqs. (\ref{eq::opt-prob-i} and \ref{eq::opt-prob-ii}) with no linear constraints. 
Attempting to include separate intercepts for overall graph and model coefficients, $\alpha_0$ and $\beta_0$ respectively, results in confounding, i.e. only $\alpha_0+\beta_0$ is estimable but they are not estimable separately. As an alternative, considering a sufficiently large graph (i.e. a graph with large number of edges $L$), one could include a group-wise intercept for prescribed groups of edges in the graph, for example, all {\em strongly connected components}. Given the nature of ${\bf W_0}$ we can penalize intercepts for neighboring groups to behave similarly, coupled with a ridge penalization to control their magnitude. However, rather than solving the estimability issue, this reduces the extent of it by confounding the intercept for chosen baseline group with true model intercept, $\beta_0$. %However, under a simple rank 1 linear constraint ${\bf 1}\trans\balpha=0$, proposed solutions produce estimates with a {\em zero-sum constraint} on the graph, that insulates the overall intercept against bias resulting from possible ``spillover" effect.  
 
In the ensuing discussion we start by deriving {\em necessary} conditions required for link functions $g_1$ and $g_2$ to satisfy algorithm (\ref{algo::md-1}), then we proceed to provide examples for different members of the Tweedie ED family under commonly used link functions. To maintain brevity we postpone necessary details to Appendix A. Unless explicitly stated, the link functions are general and not necessarily canonical. Observe that existence of a global minima is ensured by convexity of the likelihood function for member in question under specified link functions. Also, the chosen penalty function is convex, since ${\lambda_1{\bf I}_0+\lambda_2{\bf W}_0}$ is positive-semi definite for $\lambda_1, \lambda_2>0$. The conditions for theorem (\ref{th::link-fn}) remain fairly general, i.e. commonly used link functions, both canonical and general satisfy these necessary conditions. Next, we present illustrations for members of Tweedie ED family (listed in table (\ref{tab::tweedie-pars})).
\begin{mythe}\label{th::link-fn}
	If the negative log-likelihood $\ell(\Theta,p)$ is convex under chosen link functions $g_i$, with inverses $h_i$, $i=1,2$, which are
	\begin{itemize}
		\item [(i)] monotonic and invertible, implying  $h_1$ and $h_2$ exist,
		\item [(ii)] smooth inverses, implying $h_1$ and $h_2$ are at least twice continuously differentiable,
	\end{itemize}
 with $h_2\ne 0$ and, terms $\nabla_1(\bleta|\bgamma)$, $\nabla_2(\bleta|\bgamma)$, $\nabla_1(\bgamma|\bleta)$ and $\nabla_2(\bgamma|\bleta)$ in eqs. (\ref{eq::neg-lik-deriv}a--e) are well-defined in a neighborhood of the minima for objective function $F(\Theta,p)$, proposed algorithm (\ref{algo::md-1}) produces estimates that satisfy conclusions of theorem (\ref{th::conv}).
\end{mythe}

\begin{myrem}
Convexity of the negative log-likelihood ensures uniqueness of $\widehat{\Theta}$ and $p$, given that proposed penalty is convex.
\end{myrem}
\begin{myrem}
For simplicity one could assume that $h_1$ and $h_2$ are cumulative distribution functions of some random variable. In that case it is sufficient to have a differentiable probability density.
\end{myrem}

\subsection{Illustrative Examples}

Algorithm \ref{algo::md-1} is applicable to all listed members (see table (\ref{tab::tweedie-pars})) of the Tweedie ED family. Referring to the theorem (\ref{th::link-fn}) (see proof in Appendix A), quantities necessary for specifying the algorithm are, $\nabla_1(\bleta|\bgamma)$, $\nabla_2(\bleta|\bgamma)$, $\nabla_1(\bgamma|\bleta)$ and $\nabla_2(\bgamma|\bleta)$. However, while providing details for different members of the Tweedie ED family under applicable link functions $g_1$ and $g_2$, examining the existence and specifying $D(\bleta)$, $D'(\bleta)$, $D''(\bleta)$ and $a(\bgamma)$, $a'(\bgamma)$, $a''(\bgamma)$ should be sufficient.\\

\noindent {\em 1. Normal:} The normal or Gaussian distribution is a member of Tweedie ED family with index parameter $p=0$. From table (\ref{tab::tweedie-pars}), $\kappa(\theta)=\theta^2/2$, $\theta=\mu$, $\phi=\sigma^2$ and $a(y,\phi)=(2\pi\phi)^{-1/2}$ resulting in the likelihood, $f(y;\theta,\phi,p)|_{p=0}=(2\pi\phi)^{-1/2}\exp\{\phi^{-1}(y\theta-\theta^2/2)\}$. Commonly, link functions $g_1$ for $\theta=\mu$ is taken to be identity, i.e. $g_1(\mu)=\mu$ which is also the canonical link and for dispersion $\phi=\sigma^2$, $g_2(\phi)=\log(\phi)$. It follows that $h_2=\exp$, $D(\bleta)=y(x\trans\bbeta+r\trans\balpha)-(x\trans\bbeta+r\trans\balpha)^2/2$, $D'(\bleta)=y-(x\trans\bbeta+r\trans\balpha)$ and $D''(\bleta)=-1$. Additionally, $a'(\bgamma)=-\mfrac{a(\bgamma)}{2}$ and $a''(\bgamma)=\mfrac{a(\bgamma)^2}{4}$.\\
 
\noindent {\em 2. Poisson:} Poisson densities inherently feature a constant dispersion, as seen from table (\ref{tab::tweedie-pars}) and are obtained by setting $p=1$. Also, for Poisson, $\kappa(\theta)=\exp(\theta)$, $\theta=\log(\mu)$, $\phi=1$ and $a(y,\phi)=(y!)^{-1}$ resulting in the likelihood, $f(y;\theta,\phi,p)|_{p=1}=(y!)^{-1}\exp\{y\theta-\exp(\theta)\}$. The canonical link function is $g_1=\log$, $g_2=1$ however, other commonly used choices are $g_1(\mu)=\sqrt{\mu}$ and $g_1(\mu)=\mu$. Therefore, $h_1=\exp$ or $(\cdot)^2$ or identity respectively based on listed choices, while $g_2(\phi)=1$. Based on choice of the link function, (a) $g_1(\mu)=\log(\mu)$ $\implies D'(\bleta)=\big\{y-e^{x\trans\small\bbeta+r\trans\small\balpha}\big\}$, $ D''(\bleta)=-e^{x\trans\small\bbeta+r\trans\small\balpha}$, (b) $g_1(\mu)=\sqrt{\mu}\implies  D'(\bleta)=2\big\{y(x\trans\bbeta+r\trans\balpha)^{-1}-(x\trans\bbeta+r\trans\balpha)\big\}$, $D''(\bleta)=-2\big\{y(x\trans\bbeta+r\trans\balpha)^{-2}+1\big\}$, (c) $g_1(\mu)=\mu \implies  D'(\bleta)=\big\{y(x\trans\bbeta+r\trans\balpha)^{-1}-1\big\}$, $ D''(\bleta)=-y(x\trans\bbeta+r\trans\balpha)^{-2}$. Listing $D(\bleta)$ and $a(\bgamma)$ (or its derivatives) are not applicable since dispersion is constant intrinsically for Poisson densities.\\

\noindent {\em 3. Compound Poisson-gamma:} CP-g densities are characterized by lack of closed-form for $a(y,\phi)$ and feature an index parameter $1<p<2$. Lack of a closed-form expression for $a(y,\phi)$ requires approximations which we discuss in a later section. From table (\ref{tab::tweedie-pars}) note that $\theta=\mu^{(1-p)}/(1-p)$, $\kappa(\theta)=\{(1-p)\theta\}^{\frac{2-p}{1-p}}/(2-p)$ and $\phi=\phi$. Generally, $g_1=g_2=\log$, implying $h_1=h_2=\exp$, under this the negative log-likelihood is given by eqs. (\ref{eq::cpg-neg-log-lik-bessel-i}) and (\ref{eq::cpg-neg-log-lik-bessel-ii}) under chosen method of approximation for $a(y,\phi)$. Hence, $D'(\bleta)=y\exp\{(1-p)(x\trans{\small \bbeta}+r\trans{\small \balpha})\}-\exp\{(2-p)(x\trans{\small \bbeta}+r\trans{\small \balpha})\}$, $D''(\bleta)=(1-p)y\exp\{(1-p)(x\trans{\small \bbeta}+r\trans{\small \balpha})\}-(2-p)\exp\{(2-p)(x\trans{\small \bbeta}+r\trans{\small \balpha})\}$ and, $D(\bleta) = y\exp\{(1-p)(x\trans{\small \bbeta}+r\trans{\small \balpha})\}/(1-p)-\exp\{(2-p)(x\trans{\small \bbeta}+r\trans{\small \balpha})\}/(2-p)$.\\

\noindent {\em 3.1 Probability density approximations:} Approximations involving closed-form expressions for the normalizing factor $a(y,\phi,p)$ are based on (a) generalized Bessel functions, and (b) saddle-point approximations. Following paragraphs discuss necessary details for each approximation.\\

\noindent {\em 3.1.1 Generalized Bessel functions:} In CP-g densities marginalizing out the random Poisson count and equating it with eq. (\ref{eq::edm-dens}) we have $a(y,\phi,p)=y^{-1}\sum_{k=1}^{\infty}a_k(y,p)\phi^{-k(1+\xi)}I(y>0)+1\cdot I(y=0)$ as the normalizing constant associated with a particular $y$ which is a generalized Bessel function. Under a general link $\phi=h_2(z\trans\small\bgamma)$, the following result provides an approximation to relevant quantities required for algorithm (\ref{algo::md-1}).
\begin{mylem} 
	If $g^{-1}_2= h_2$ satisfies conditions of theorem (\ref{th::link-fn}), then summands in expressions for $a(\bgamma)$, $a'(\bgamma)$ and $a''(\bgamma)$ as generalized Bessel functions, have approximate modes at 
	\begin{align*}
	k_{\max}=\frac{y^{2-p}}{(2-p)h_2(z\trans\small\bgamma)}.
	\end{align*}
\end{mylem}

\noindent {\em 3.1.2 Saddlepoint approximation:} Expressed in an {unit-deviance form}, the negative log likelihood for a CP-g density is given by eqs. (\ref{eq::cpg-neg-log-lik-saddle-i}) and (\ref{eq::cpg-neg-log-lik-saddle-ii}) under $g_1=g_2=\log$. Using $b(\bgamma)=\big(2\pi V(y) h_2(z\trans\bgamma)\big)^{-1/2}=\big(2\pi y^p e^{z\trans\bgamma}\big)^{-1/2}\cdot I(y>0)+\big(2\pi \epsilon_0^p e^{z\trans\bgamma}\big)^{-1/2}\cdot I(y=0)$ produces a saddlepoint approximation to CP-g density. The deviance is
\begin{align}\label{eq::tw-dev-i}
	d(y,\bleta)=2\Bigg\{y\frac{y^{1-p}-e^{(1-p)(x\trans{\small\bbeta}+r\trans{\small\balpha})}}{1-p}-\frac{y^{2-p}-e^{(2-p)(x\trans{\small\bbeta}+r\trans{\small\balpha})}}{2-p}\Bigg\}.
\end{align} 
If $y=0$, $d(y,\bleta)=\exp\{(2-p)(x\trans\bbeta+r\trans\balpha)\}\big/(2-p)$, furthermore, noticing that $d(y,\bleta)=D(y)-D(\bleta)$, $b'(\bgamma)=-\mfrac{b(\bgamma)}{2}$ and $b''(\bgamma)=\mfrac{b^2(\bgamma)}{4}$ specifications for algorithm (\ref{algo::md-1}) are complete. $\epsilon_0$ is chosen to be optimally small. \\

\noindent {\em 3.2 Estimation of index parameter:} Based on the relationship $\text{Var}(y)=\phi V(\mu)=\phi\mu^p$ in eq. (\ref{eq::graph-DGLM}), an empirical estimate of $p$ is obtained by plotting log-variance against log-mean across multiple splits of the data. However, a model based criteria is obtained by maximizing the {\em profile likelihood} associated with $p$. This agrees with the inherent definition of $p$ for Tweedie EDs, that is related to model selection.  Analytically, this translates to maximizing the likelihood for a line or grid of values chosen for the index parameter $p$ such that $1<p<2$. Proposed algorithm (\ref{algo::md-1}) performs automatic model selection by updating $p$ using the profile likelihood.\\

\noindent {\em 4. Gamma:} Gamma densities are characterized by a positive continuous real support. Within Tweedie ED models they are specified by an index parameter, $p=2$. We have $\theta=-(\mu)^{-1}$, $\kappa(\theta)=-\log(-\theta)$, $\phi=\phi$ and $a(y,\phi)=\phi^{-1/\phi}y^{1/\phi-1}\big(\Gamma(1/\phi)\big)^{-1}$ resulting in the likelihood, $f(y;\theta,\phi,p)|_{p=2}=\phi^{-1/\phi}y^{1/\phi-1}\big(\Gamma(1/\phi)\big)^{-1}\exp\big\{\phi^{-1}(y\theta+\log(-\theta))\big\}$. Listing commonly used link functions and relevant quantities,
 (a) $g_1(\mu)=1/\mu \implies$ $D(\bleta)=-y(x\trans\bbeta+r\trans\balpha)+\log(x\trans\bbeta+r\trans\balpha)$, $D'(\bleta)=-y+(x\trans\bbeta+r\trans\balpha)^{-1}$, $D''(\bleta)=-(x\trans\bbeta+r\trans\balpha)^{-2}$,
  (b) $g_1(\mu)= \mu \implies$ $D(\bleta)=-y(x\trans\bbeta+r\trans\balpha)^{-1}-\log(x\trans\bbeta+r\trans\balpha)$, $D'(\bleta)=y(x\trans\bbeta+r\trans\balpha)^{-2}-(x\trans\bbeta+r\trans\balpha)^{-1}$, $D''(\bleta)=-2y(x\trans\bbeta+r\trans\balpha)^{-3}+(x\trans\bbeta+r\trans\balpha)^{-2}$ and,
   (c) $g_1(\mu)=\log(\mu)\implies$ $D(\bleta)=-ye^{-(x\trans\small\bbeta+r\trans\small\balpha)}-(x\trans\bbeta+r\trans\balpha)$, $D'(\bleta)=ye^{-(x\trans\small\bbeta+r\trans\small\balpha)}-1$, $D''(\bleta)=-ye^{-(x\trans\small\bbeta+r\trans\small\balpha)}$. 
   Additionally, under $g_2(\phi)=\log(\phi)$ we have $a(\bgamma)=\exp(-e^{-z\trans\small\bgamma}z\trans\bgamma)y^{e^{-z\trans\small\bgamma}-1}\big[\Gamma(e^{-z\trans\small\bgamma})\big]^{-1}$. Note that, if $\psi^{(m)}(\cdot)$ denotes the {\em polygamma} function of order $m$, i.e. $\psi^{(m)}(t)=\frac{d^{m+1}}{dt^{m+1}}\log \Gamma(t)$,
   \begin{align*}
   	\log(a(\bgamma))&=-e^{-z\trans\small\bgamma}z\trans\bgamma+(e^{-z\trans\small\bgamma}-1)\log(y)-\log\big[\Gamma(e^{-z\trans\small\bgamma})\big],\\
   a'(\bgamma)&=a(\bgamma)\big[-e^{-z\trans\small\bgamma}+e^{-z\trans\small\bgamma}z\trans\bgamma-e^{-z\trans\small\bgamma}\log(y)+\psi^{(0)}\big(\Gamma(e^{-z\trans\small\bgamma})\big)\big],\\
    a''(\bgamma)&=a(\bgamma)\big[e^{-z\trans\small\bgamma}z\trans\bgamma+e^{-z\trans\small\bgamma}\log(y)-\psi^{(1)}\big(\Gamma(e^{-z\trans\small\bgamma})\big)\big]+a'^2(\bgamma)a(\bgamma)^{-1},
   \end{align*}                  
 	which completes the specification. Polygamma functions are analytically tractable and provided by most commonly used computational softwares. Furthermore, using saddlepoint approximations provide simpler and effective analytic expressions.\\
 	
\noindent {\em 5. Inverse Gaussian:} Inverse Gaussian densities also have a positive support with an index parameter $p=3$. Link functions used are $g_1(\mu)=\mu^{-2}$, $g_2(\phi)=\log(\phi)$. We have $\theta=-(2\mu)^{-2}$, $\kappa(\theta)=-\sqrt{-2\theta}$, $\phi=\phi$ and $a(y,\phi)=(2\pi y^3\phi)^{-1/2}$ resulting in the likelihood $f(y;\theta,\phi,p)|_{p=3}=(2\pi y^3\phi)^{-1/2}\exp\{\phi^{-1}(y\theta+\sqrt{-2\theta})\}$. Under chosen links $D(\bleta)=-2y(x\trans\bbeta+r\trans\balpha)+(x\trans\bbeta+r\trans\balpha)^{1/2}$, $D'(\bleta)=-2y+\frac{1}{2}(x\trans\bbeta+r\trans\balpha)^{-1/2}$ and, $D''(\bleta)=-\frac{1}{4}(x\trans\bbeta+r\trans\balpha)^{-3/2}$. Under $g_2(\phi)=\log(\phi)$, $a'(\bgamma)=-\mfrac{a(\bgamma)}{2}$ and $a''(\bgamma)=\mfrac{a(\bgamma)^2}{4}$.     

\subsection{Scalability via approximation}

Scalability of algorithm (\ref{algo::md-1}) is affected adversely by multiple factors viz., large $k_\beta$ or $k_\gamma$ or both, large $N$ and large $L$. Attributing first two problems to the {\em curse of dimensionality}, we attempt to construct an approximate algorithm for graphs featuring large $L$.  Referring to eq. (\ref{eq::neg-lik-deriv}b) and (\ref{eq::sol-eta})  and noting that the matrix $\nabla_2(\bleta|\bgamma)$ is defined as a partitioned matrix we have 
\begin{align*}
	\lambda_1{\bf I}_0+\lambda_2{\bf W}_0+c_1\nabla_2(\bleta|\bgamma)=\arraycolsep=3.0pt\def\arraystretch{1.5}\left[\begin{array}{c:c} c_1\nabla_2^{(11)} & c_1\nabla_2^{(12)}\\\hdashline c_1\nabla_2^{(21)} & \lambda_1I_L+\lambda_2\mathfrak{W}+c_1\nabla_2^{(22)} \end{array}\right],
\end{align*}
\begin{align*}
	\left[\lambda_1{\bf I}_0+\lambda_2{\bf W}_0+c_1\nabla_2(\bleta|\bgamma)\right]^{-1}=\arraycolsep=3.0pt\def\arraystretch{1.5}\left[\begin{array}{c:c} c_1^{-1}\mathfrak{A}^{-1}& -\mathfrak{A}^{-1}\nabla_2^{(12)}\mathfrak{S}_{22}^{-1}\\\hdashline -\mathfrak{S}_{22}^{-1}\nabla_2^{(21)} \mathfrak{A}^{-1}& \mathfrak{S}_{22}^{-1}+ c_1\mathfrak{S}_{22}^{-1}\nabla_2^{(21)}\mathfrak{A}^{-1}\nabla_2^{(12)}\mathfrak{S}_{22}^{-1}\end{array}\right],
\end{align*}
 where $\lambda_1I_L+\lambda_2\mathfrak{W}+c_1\nabla_2^{(22)}= \mathfrak{S}_{22}$, $\mathfrak{A}=\nabla_2^{(11)}-\nabla_2^{(12)}\mathfrak{S}_{22}^{-1}\nabla_2^{(21)}$ and $\nabla_2^{(22)}$ is a diagonal matrix since, $r_{i'j}\trans r_{ij}=0$ if $i\ne i'$. For graphs with large number of vertices $L$, we construct approximate Laplacian matrices following Halder et al. (2019) \cite{halder2019spatial}, which results in $\mathfrak{S}_{22}$ being a block diagonal matrix, consequently improving computational efficiency for calculating $\mathfrak{S}_{22}^{-1}$ iteratively in eq. (\ref{eq::sol-eta}), thereby improving scalability of algorithm (\ref{algo::md-1}). Halder et al. (2019) compares efficiency of produced estimates under approximate and exact solutions demonstrating the trade-off between computational complexity and accuracy for CP-g DGLMs.
 
It is clear from preceding examples that within the ED family, CP-g models present fair amount of challenges due to their complex model structure. Response is therefore composed of exact zeros coupled with positive continuous observations. Observations featuring exact zeros produce no signal, naturally rendering {\em signal-to-noise ratio} (SNR) as an important parameter for judging performance. For computational feasibility and scalability we employ a {\em saddle-point approximation} for the CP-g likelihood. In what follows we present details for simulation and real data analysis with the purpose of showcasing performance of proposed algorithm for CP-g DGLMs. Noting that the nature of methods developed in this section remain fairly general, similar results can be produced for other members of the family and any undirected graph. We strictly maintain a spatial interpretation for the associated undirected graph here onward.
 
\section{Simulation}\label{sec::sim}

\begin{figure}[t]
	\centering
	\begin{subfigure}{.25\textwidth}
		\centering
		\includegraphics[width=1\linewidth , height=0.8\linewidth]{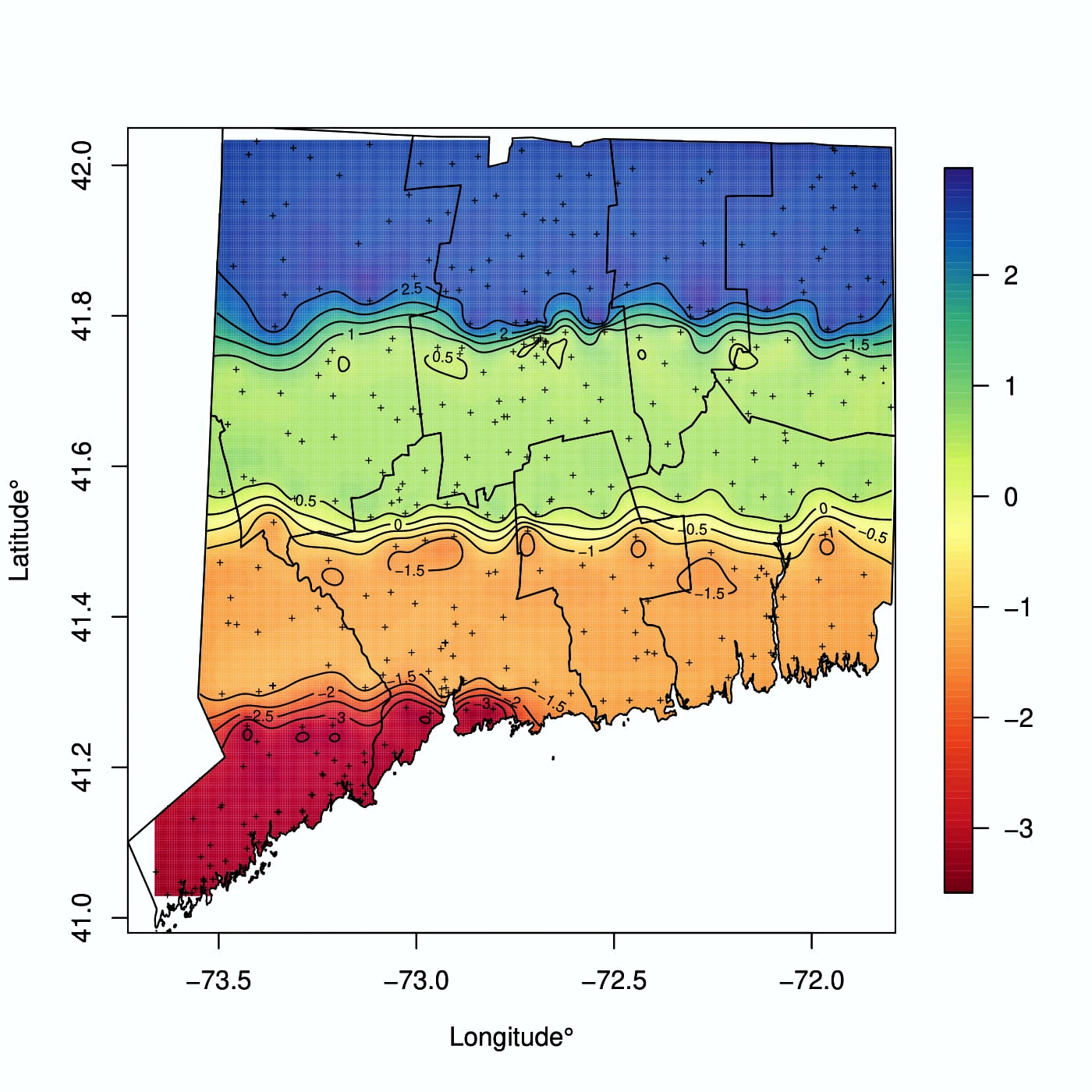}
		\caption{}
	\end{subfigure}%
	\begin{subfigure}{.25\textwidth}
		\centering
		\includegraphics[width=1\linewidth , height=0.8\linewidth]{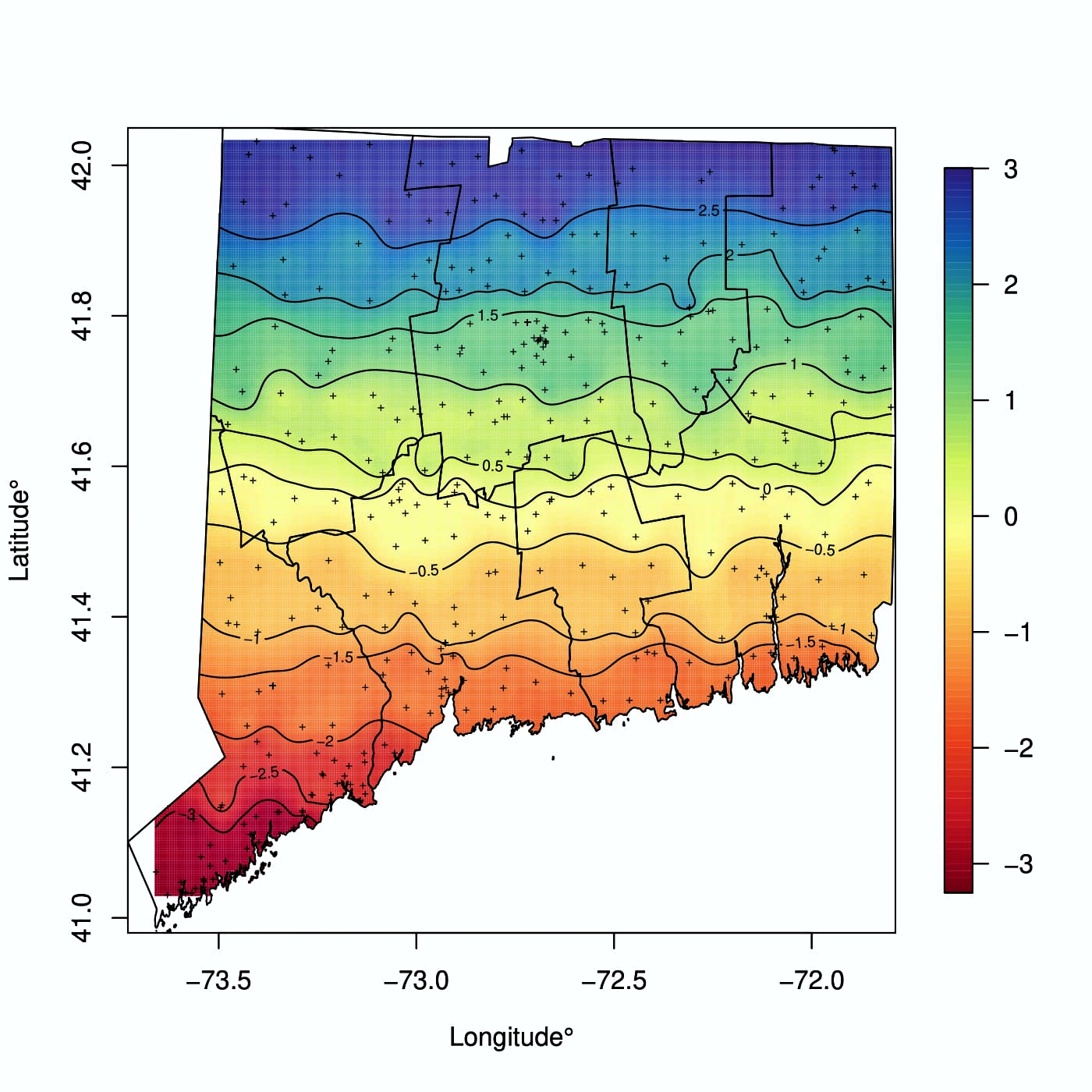}
		\caption{}
	\end{subfigure}%
	\begin{subfigure}{.25\textwidth}
		\centering
		\includegraphics[width=1\linewidth , height=0.8\linewidth]{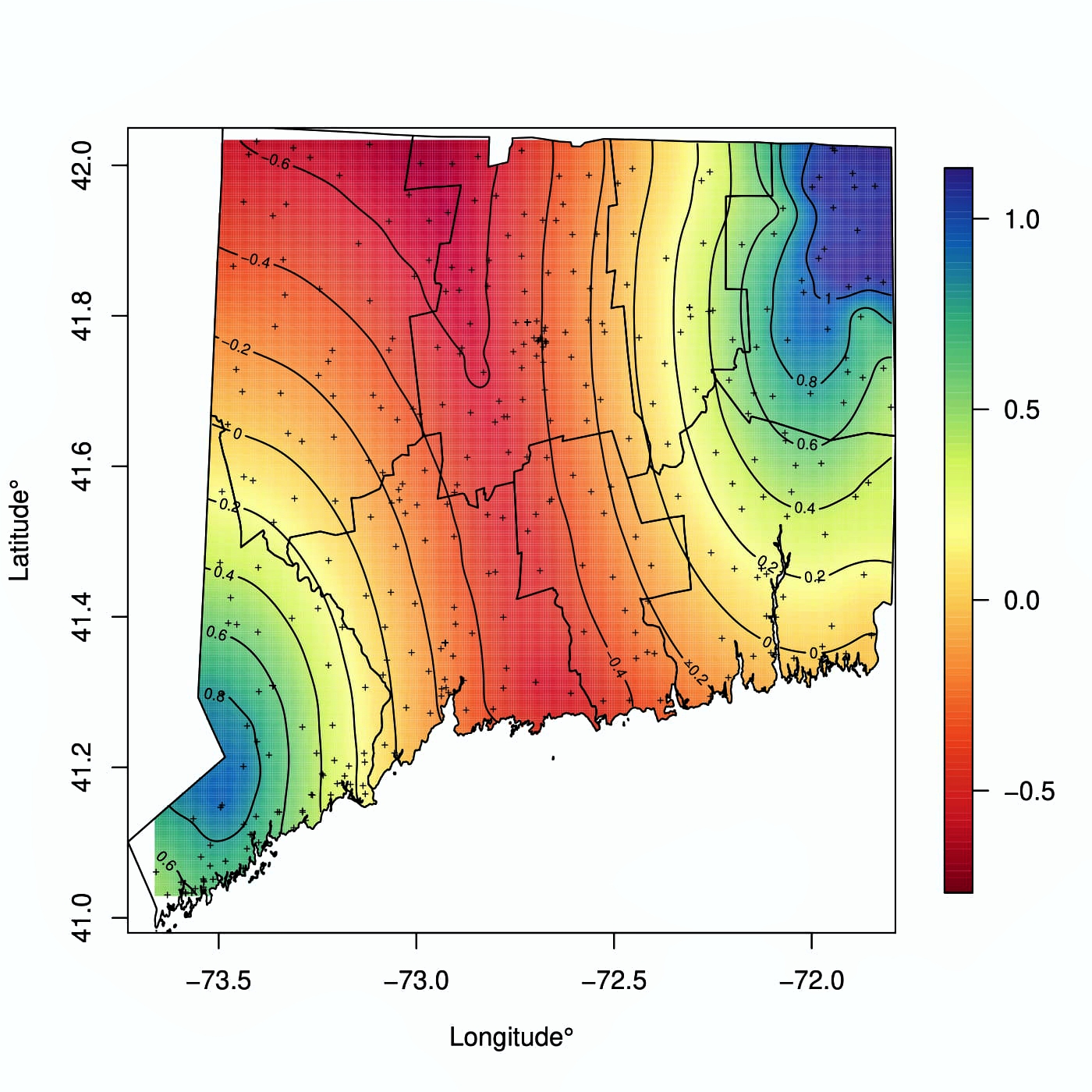}
		\caption{}
	\end{subfigure}%
	\begin{subfigure}{.25\textwidth}
		\centering
		\includegraphics[width=1\linewidth , height=0.8\linewidth]{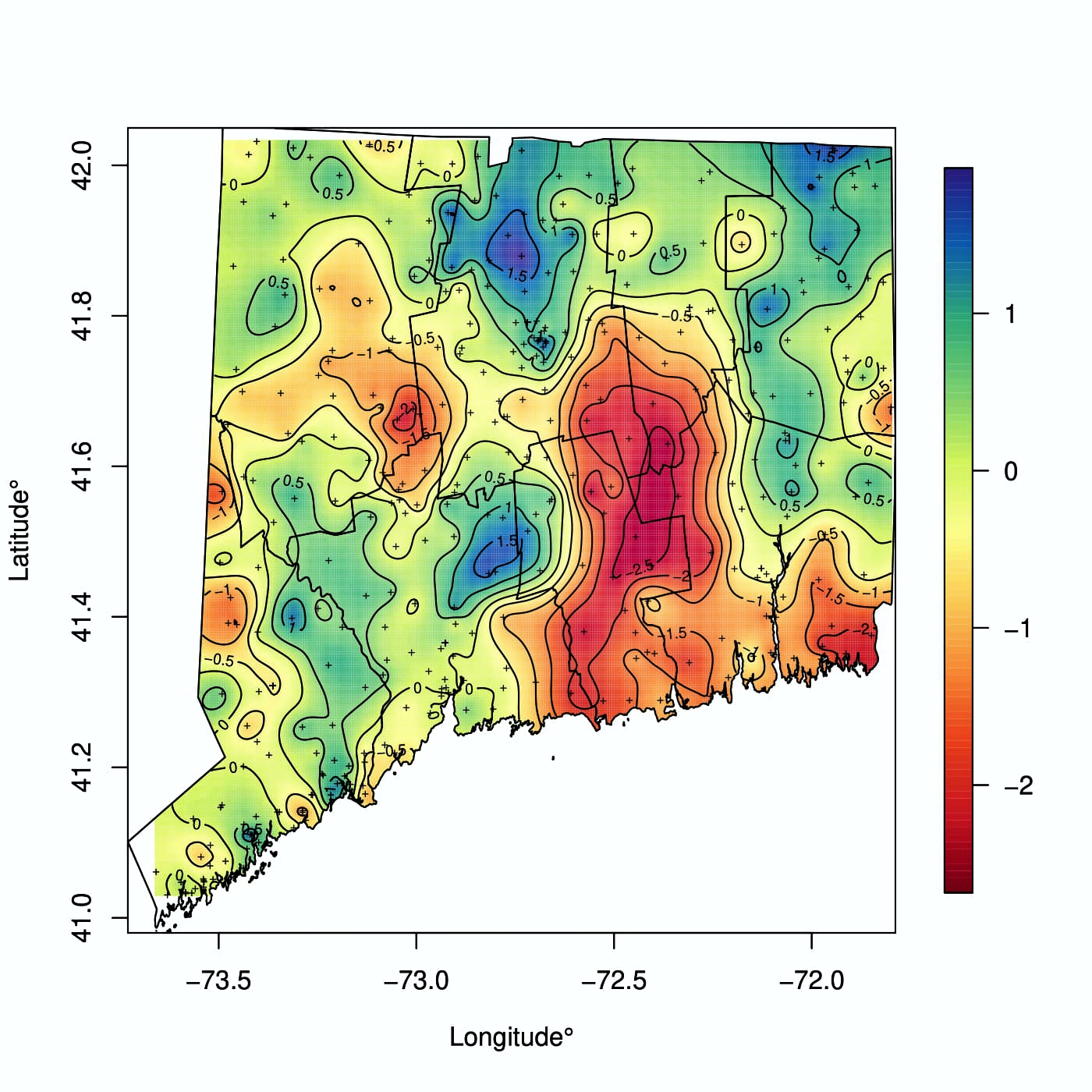}
		\caption{}
	\end{subfigure}
	\caption{Spatial patterns (a) block (horizontal), (b) smooth (horizontal), (c) hot-spots and, (d) structured.}
	\label{fig::ct-sim}
\end{figure}

In this section we aim to demonstrate the performance of proposed algorithm (\ref{algo::md-1}) under varying (a) sample sizes (b) proportion of zeros and, (c) spatial patterns. Proportion of zeros serve as an indicator for the amount of signal in synthetic response, thus quantifying an SNR parameter. Spatial patterns used to elucidate performance are shown in figure (\ref{fig::ct-sim}). In particular, the last pattern considered, that is labeled structured in fig. (\ref{fig::ct-sim}d), is composed of centered normally distributed random spatial effects, with zero mean and a squared exponentially {\em structured} covariance kernel, $\sigma^2\exp\left(-\phi||\alpha_i-\alpha_j||_2^2\right)$, $i,j=1,2,\ldots,L$, with $\sigma^2=1.5$ and $\phi=3$. The reason behind including this is to demonstrate that although suggested penalty is based out of a neighbor-based adjacency, its efficacy in detecting patterns that are distance based is not compromised. The state of Connecticut (CT) serves as a placeholder, providing a fixed spatial reference frame, relying on which we generate synthetic response. We begin by defining metrics used to compare performance of proposed estimates with appropriate alternatives viz., (i) unpenalized estimates (MLE) and (ii) ridge estimate. Before providing closed forms for these alternatives we define necessary metrics. Denote $\Theta^{(0)}$, $\widehat{\Theta}_0$, $\widehat{\Theta}_r$ and $\widehat{\Theta}$ as the oracle, unpenalized, ridge and proposed estimated parameters respectively. On the training split of a particular synthetic instance, to compare accuracy of estimated parameters we use sum-of-squares of error, $SSE(a)=||\Theta^{(0)}-a||^2_2$, where $a$ can be any one of the discussed estimates, while on the validation set, predictive accuracy is judged using a deviance ratio, $D(a)=d(y,a)\big/d(y,\bleta^{(0)})$, where the deviance, $d(y,\bleta)$ is shown in eq. (\ref{eq::tw-dev-i}). Lower $SSE(a)$ and $D(a)$ close to 1 are desirable. The MLE and ridge alternatives used for comparison have the following closed forms,
\begin{gather*}
		\widehat{\bleta}_0= \bleta-c_1^{-1}\nabla_2(\bleta|\bgamma)^{-1}\big[\nabla_1(\bleta|\bgamma)\big],\\%+{\bf B}_\eta{\bf c}_\eta
		\widehat{\bleta}_r= [\lambda_1{\bf I_0}+c_1\nabla_2(\bleta|\bgamma)]^{-1}\{c_1\nabla_2(\bleta|\bgamma)\bleta-\nabla_1(\bleta|\bgamma)\},%-{\bf B}_\eta{\bf c}_\eta
\end{gather*}
while $\widehat{\bgamma}_0=\widehat{\bgamma}_r=\widehat{\bgamma}$. Initially, we use an index parameter $p=1.5$ and $g_1=g_2=\log$. Later, we demonstrate use of profile likelihood to choose the index parameter. As mentioned, synthetic data is separated into training and validation sets, in particular we use a 60-40 split for the scope of this simulation. Tuning parameters $\lambda_1$ and $\lambda_2$ are estimated through minimizing deviance (as shown in eq. (\ref{eq::tw-dev-i})) over a hold-out set while using {\em leave-one-out cross-validation} (LOOC) on the training split. As a consequence of theorem (\ref{th::conv}) accuracy of proposed estimates are subject to the chosen grid. For the scope of this paper we optimize over $\big(\log(\lambda_1),\log(\lambda_2)\big)\in [-5,5]\times [-5,5]$ or appropriate subsets. However for a ridge solution we optimize over a line, $\log(\lambda_1)\in [-5,5]$. We generate both continuous and categorical synthetic covariates. Intercepts are included for the mean and dispersion model separately. We use four covariates $X_1\sim Bin(1,0.5)$, $X_2\sim Bin(4,0.5)$, $X_3\sim N(0,0.1))$ and $X_4\sim N(0,0.1)$ separately for the mean and dispersion models. Chosen sample sizes vary from 10,000 -- 50,000, with SNRs varying from 15\% -- 80\%.  Further details regarding the generation of synthetic response for various spatial patterns and SNRs are omitted to keep this paper concise (see section 4 of Halder et al. (2019) (\cite{halder2019spatial})). Tables (\ref{tab::sim-bl}), (\ref{tab::sim-sm}), (\ref{tab::sim-hs}) and (\ref{tab::sim-str}) in Appendix B show results of the simulation for respective patterns, sample size and SNR settings, which are summarized below.
\begin{itemize}
	\setlength\itemsep{0.01in}
	\item [(i)] Under all different spatial patterns, proposed estimates show lower sum of square for error, i.e.  $SSE(\Theta)=||\Theta^{(0)}-\Theta||_2^2=||\bbeta^{(0)}-\bbeta||_2^2+||\balpha^{(0)}-\balpha||_2^2+||\bgamma^{(0)}-\bgamma||_2^2$ is the lowest, showcasing {\em accuracy}. \item [(ii)] In particular under low SNRs (i.e. higher proportion of zeros) performance of proposed estimates $\Theta$, remain superior in comparison to the alternatives, showcasing {\em stability}.
	\item [(iii)] The standard deviation associated with $SSEs$ for proposed estimates are the lowest, showcasing {\em consistency}.
	\item [(iv)] Deviance ratios in validation sets remain desirably close to 1 for proposed estimates, showcasing its {\em ability to detect a spatial effect} (if present).
\end{itemize}

A {\em warm-start} strategy is used while performing LOOC on synthetic data. Explicitly, while updating interior grid points $(\lambda_1,\lambda_2)$, we use estimated parameters, $\{\widehat{\Theta},\widehat{p}\}$, under the immediately previous grid coordinate as a starting point for the current coordinate. Fitted estimates are obtained at an interior point featuring the minimum deviance within the grid. Finally, asymptotic inference on $\widehat{\Theta}$ proceeds on noting that, $\widehat{\Theta}\big|_{p=\widehat{p}} \stackrel{a}{\sim} \mathcal{N}(\Theta, \Sigma_{\Theta})$, where $ \Sigma_\Theta=E\big[\big(\frac{\partial^2}{\partial\Theta^2}\ell(\Theta,p)\big|\Theta\big)^2\big]^{-1}$ is the associated Fisher's information matrix. Following this we can compute significance for fitted fixed effects using approximate $p$-values based on the asymptotic distribution for $\widehat{\Theta}$. 

\section{Real Data}\label{sec::real-apply}

Proposed algorithm is applied to a real data instance pertaining to the state of Connecticut. The response is chosen to be {\em loss costs per unit insured}, which is derived as the ratio of incurred losses to policy exposure, under automobile collision coverage for policies from a chosen cohort of insurance companies in the year 2008. The data is obtained from a much more comprehensive repository named Highway Loss Data Institute (HLDI) maintained by the independent non-profit, Insurance Institute for Highway Safety (IIHS) \cite{iihs} working towards reducing losses arising from motor vehicle collisions in North America. We shall refer to this as the HLDI database.

We briefly describe the HLDI database. It contains data at an individual level. The data contains covariates associated with the individual at,
\begin{itemize}
	\item [a.] {\bf Individual Level}
	\begin{enumerate}
		\item accident and model year of the vehicle, ranging from 2000--2015 and 1981 -- 2016 respectively,
		\item risk having two levels ``S", ``N",
		\item age, gender, marital status and gender of partner, where missing values are denoted by 0 and ``U" respectively for age and rest of these predictors,
	\end{enumerate}
	\item [b. ] {\bf Policy level}
	\begin{enumerate}
		\item number of claims accompanied by associated payments (i.e. loss cost),
		\item exposure (measured in policy years, eg. 0.5 indicates individual insured for half a year)
		\item deductible limit (categorical with 8 categories)
	\end{enumerate}
	\item[c. ] {\bf Spatial:} 5-digit zip code indicating location, i.e. areally-referenced.
\end{itemize} 
Derived predictors like age categories, vehicle age in years are computed and used while fitting models. The last level (for ex.  unspecified, or missing) is generally omitted to create dummy variables. The number of records for the state of Connecticut was 1,513,655 (approx. 1.5 million) for chosen year 2008, with 282 zip codes. The proportion of zeros at the observation level was 95.73\%. We use a 60--40 training and validation split. The average out-sample policy premium was 183.90 US Dollars.

\begin{figure}[t]
  	\centering
  	\begin{subfigure}{.33\textwidth}
  		\centering
  		\includegraphics[width=1\linewidth , height=0.9\linewidth]{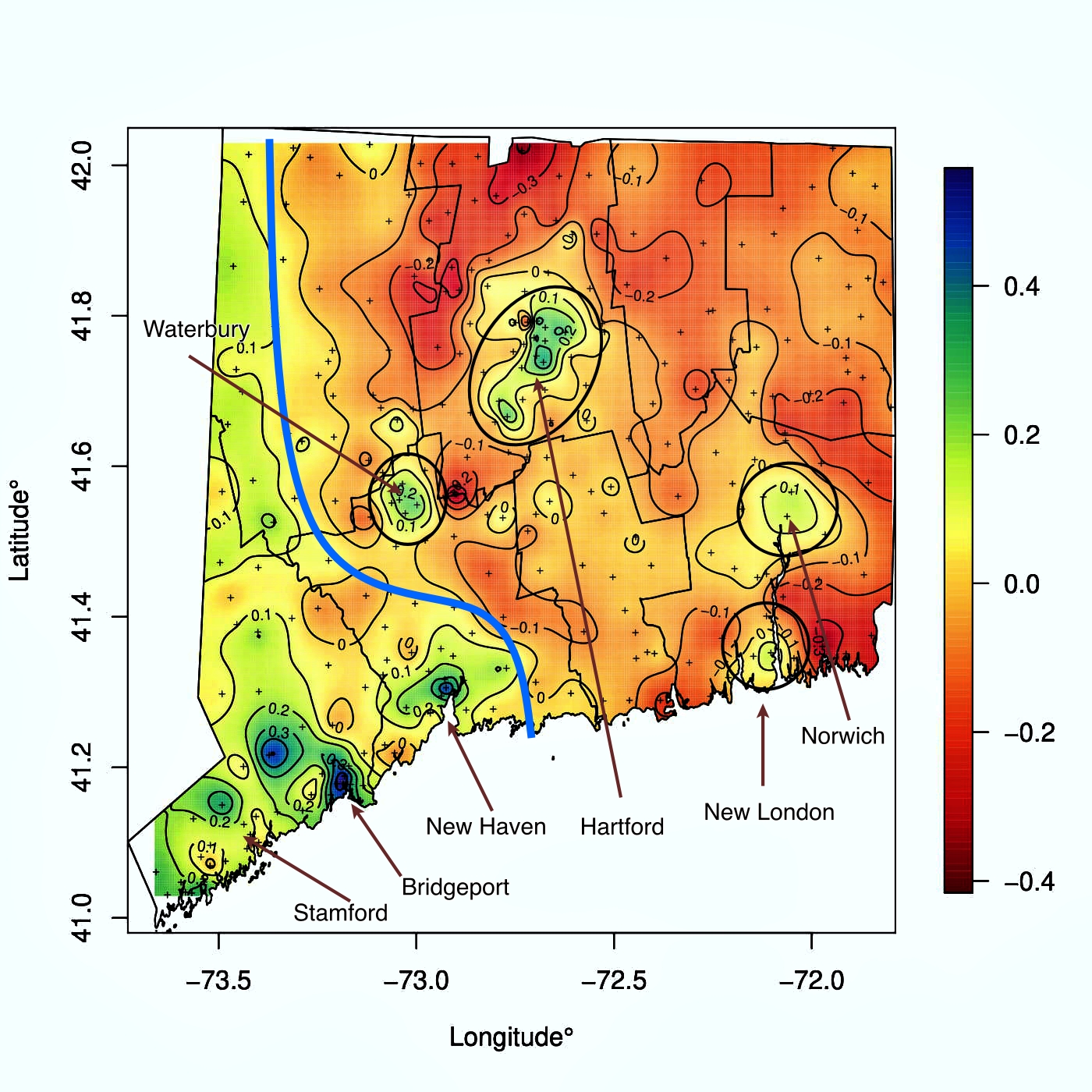}
  		\caption{}
  	\end{subfigure}%
  	\begin{subfigure}{.33\textwidth}
  		\centering
  		\includegraphics[width=1\linewidth , height=0.9\linewidth]{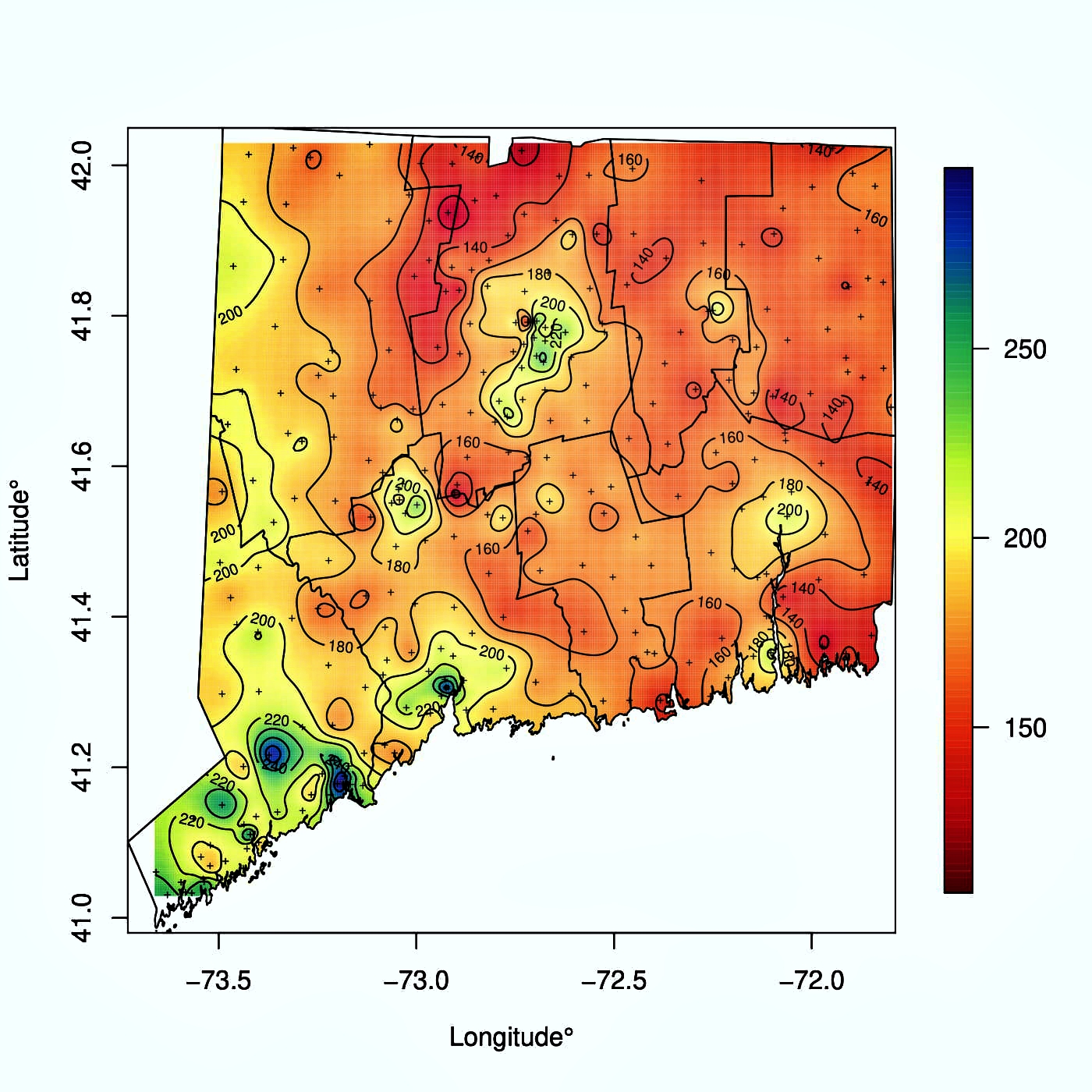}
  		\caption{}
  	\end{subfigure}%
  	\begin{subfigure}{.33\textwidth}
  		\centering
  		\includegraphics[width=1\linewidth , height=0.9\linewidth]{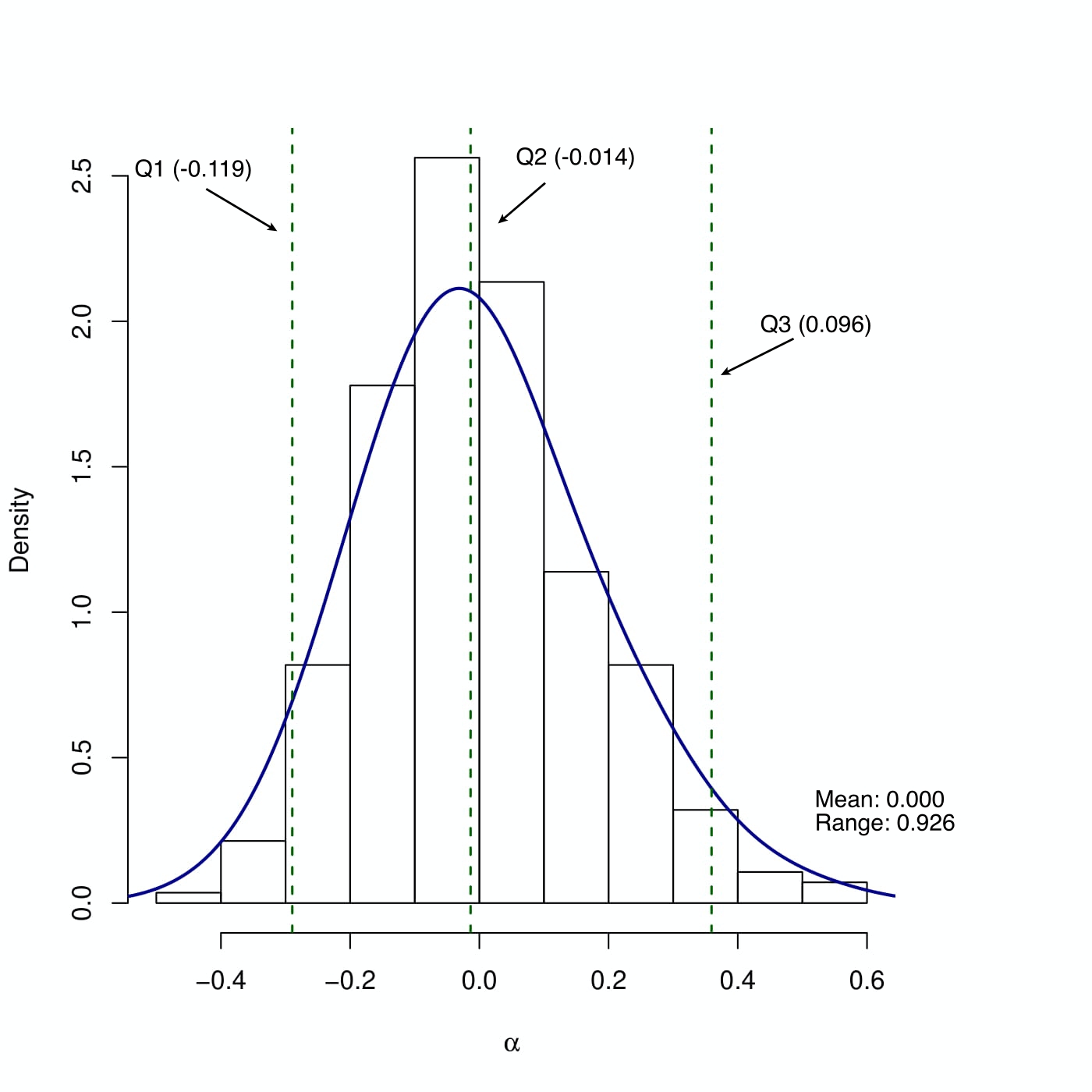}
  		\caption{}
  	\end{subfigure}%
  	\caption{Krigged spatial plots showing (a) estimated effects $\widehat{\balpha}$, marked with zones delineating separating boundaries between positive and negative spatial effects (marked using a blue line). Prominent cities are circled. (b) fitted mean policy premium (in US dollars) at the zip code level. The map shows 8 county borders and 282 zip codes marked using ``+". Figure (c) shows a histogram for estimated spatial effects with necessary numerical summarizations.}
  	\label{fig::ct-real}
  \end{figure}

To adjust for policy exposure (duration), we use the scale invariance property of Tweedie densities, i.e. if policy exposure is denoted by $w$, $\phi^*=\phi/w$ and, $y^*=y/w$ then $y^*\sim Tw(\mu,\phi^*,p)$. The fitted spatial DGLM is as follows

\begin{align*}
	\log(\mu)=\beta_0+\beta_1 \texttt{Vehicle Age}+\beta_2 \texttt{Risk}+\beta_3 \texttt{Age Class}+\beta_4 \texttt{Gender}+\beta_5 \texttt{Marital Status}\\+\beta_6 \texttt{Deductible Class}+\beta_7 \texttt{Marital Gender}+r\trans\balpha,
\end{align*}

\begin{align*}
\log(\phi^*)=\gamma_0+\gamma_1 \texttt{Vehicle Age}+\gamma_2 \texttt{Risk}+\gamma_3 \texttt{Age Class}+\gamma_4 \texttt{Gender}+\gamma_5 \texttt{Marital Status}\\+\gamma_6 \texttt{Deductible Class}+\gamma_7 \texttt{Marital Gender},
\end{align*}
where $\balpha$ is the $282\times 1$ unobserved spatial effect vector and, $r\trans$ is the observation level membership vector with respect to zip code. Tuning parameters $\log(\lambda_1)$, $\log(\lambda_2)$ are estimated by minimizing out-sample deviance,
\begin{align*}
		d\left(y^*,\bleta\big|w\right)=2w\left\{y^*\frac{{y^*}^{1-p}-e^{(1-p)(x\trans{\small\bbeta}+r\trans{\small\balpha})}}{1-p}-\frac{{y^*}^{2-p}-e^{(2-p)(x\trans{\small\bbeta}+r\trans{\small\balpha})}}{2-p}\right\},
\end{align*}
  on the holdout split while using LOOC over a $20\times 20$ grid within $[-5,5]\times [-5,5]$, with optimal values at $\lambda_1=2.20$, and $\lambda_2=30.60$. Here $y$ is the response, i.e. loss incurred per unit exposure, $w$. The index parameter was estimated at,  $\widehat{p}=1.8$. Resulting estimates $\widehat{\bbeta}$ and $\widehat{\bgamma}$ are shown in tables (\ref{tab::mean-mod} and \ref{tab::disp-mod}). Estimated spatial effects and fitted aggregated zip-code level mean policy premiums (in US dollars) are shown in figures (\ref{fig::ct-real}a, \ref{fig::ct-real}b). Figure (\ref{fig::ct-real}c) shows a histogram with the kernel density estimate for $\widehat{\balpha}$, the median $\tilde{\balpha}=-0.014$, mean $\bar{\balpha}=0.000$, standard deviation of $0.170$ and range of $0.926$. Consequently, the overall spatial effect exerts no influence on the overall mean, but accounts for introducing over-dispersion in the fitted CP-g DGLM. The predicted mean out-sample policy premium was 186.70 US dollars. Fitted dispersion $\phi^* \in (9.40, 53.35)$, while predicted dispersion ranged from $(8.38,53.35)$.
  
  From tables (\ref{tab::mean-mod}) and (\ref{tab::disp-mod}), we conclude that marital status and gender of spouse are both not significant for mean and dispersion models. Comparing effects for both models we notice that the nature of effect for a variable differs based on the model. For instance, \texttt{Vehicle Age} shows significant positive effects for the mean model, however they are significant negative in the dispersion model. This is again noticed for different levels of \texttt{Deductible Class} comparing the mean and dispersion models. However, variables like \texttt{Age Class} and \texttt{Risk} exert the same effect in both models. 
  
  Comparing estimated zip code level spatial effects $\widehat{\balpha}$ across the state, from figure (\ref{fig::ct-real}a) we see that big cities are generally {\em densely populated} and therefore accompanied by a relatively large positive effect as opposed to sparsely populated areas in the north eastern part. Furthermore, we see a clear separating boundary within the state, separating large zones of positive and negative spatial effects. The western part of the state has denser population accompanied by a densely populated neighboring city/state New York, New York as opposed to its eastern neighbor, Rhode Island, which explains the gradient in $\widehat{\balpha}$. Lastly, the monthly premium (shown in figure (\ref{fig::ct-real})b) shows a similar gradient when compared to $\widehat{\balpha}$ which is intuitive referring to the model in eq. (\ref{eq::graph-DGLM}) under a logarithm link.

\section{Remarks}\label{sec::remarks}

Developments regarding proposed framework in the paper produces numerous avenues to be pursued in future research. In the ensuing discussion, we conclude our efforts by mentioning milestones established and discuss possible future research. We also discuss future availability of computational software for the purpose of dissemination of proposed methods, which include necessary sub-routines for fitting DGLMs, synthetic data generators and example real data instances for open-source access.

\subsection{Conclusion}

 We have developed a co-ordinate descent algorithm for fitting exponential dispersion Tweedie double generalized linear models, while simultaneously performing spatial uncertainty quantification. DGLMs provide a more general framework when compared to GLMs, that allow for modeling of dispersion. Furthermore, spatial effects are generally unobserved, therefore proposed model is asymptomatic to including random effects in model (\ref{eq::DGLM}). We produced a convergence proof in theorem (\ref{th::conv}) which guarantees an accuracy of $2\epsilon/\lambda_1$, $\lambda_1>0$, in the Euclidean metric, where chosen $\epsilon$ can be made arbitrarily small, in proposed estimates. Extensive simulations showcase superiority and efficacy of proposed estimates over available state of the art methods under a variety of sample sizes and signal-to-noise ratios. Scalability is approached via approximate adjacency under the assumption that underlying latent unobserved surface is smooth. Developed framework is then used to model a real data instance for automobile collision insurance coverage in the state of Connecticut, 2008. Produced fixed effects $\widehat{\bbeta}$, $\widehat{\bgamma}$ for the mean and dispersion model respectively provides insight into underlying dependence of the response, pure-premium on the covariates used in respective models. Produced spatial effects show similarity to {\em population density}, which is intuitive given the nature of response.
  
\subsection{Discussion and Future Work}

An \texttt{R}-package (\cite{rcite}) containing the full implementation for all members of the family including the approach in (\cite{halder2019spatial}) is being created. The nature of developed methods remain fairly general and applicable to all members in the Tweedie ED family. Proposed penalty can be extended to have a ridge penalization for $\widehat{\bbeta}$ and $\widehat{\bgamma}$, which is straightforward. In future extending these developments to include temporal modeling would provide further generalization to models shown in this paper. Furthermore, within a DGLM framework the nature of spatial dependence for mean and dispersion models can be different. Using similar methodology we can provide spatial uncertainty quantification for the dispersion model. Following such pursuits, we aim to look into a penalty that produces simultaneous model selection (variable selection). Since spatial dependence adjust for bias in estimated fixed effects, variables selected should differ in mean and dispersion models respectively providing a much need procedure for model specification. Naturally, such pursuits should not be restricted to a grid-based optimality constraint. Finally, as ambitious as the stated aims are, we feel that such pursuits should provide a flexible framework for the Tweedie ED family.

\newpage

\section*{Appendix A: Proofs}
\noindent {\em Proof of Theorem 1:}\\

\noindent By definition $F(\Theta^*,p^*)\leq F(\Theta^*,p)$, consequently the first inequality is true. Define, $\mathcal{L}_p(\bleta^*|\bleta,\bgamma)=\ell(\Theta,p)+(\bleta^*-\bleta)\trans\nabla_1(\bleta|\bgamma)+\mfrac{1}{2}(\bleta^*-\bleta)\trans c_1\nabla_2(\bleta|\bgamma)(\bleta^*-\bleta)$ then for any $\bdelta \in \mathbb{R}^{k_\beta+L}$,
\begin{eqnarray}
	\mathcal{L}_p(\bleta^*+\bdelta|\bleta,\bgamma)-\mathcal{L}_p(\bleta^*|\bleta,\bgamma)&=&\scalemath{0.9}{\ell(\Theta,p)+(\bleta^*+\bdelta-\bleta)\trans\nabla_1(\bleta|\bgamma)+\mfrac{1}{2}(\bleta^*+\bdelta-\bleta)\trans c_1\nabla_2(\bleta|\bgamma)(\bleta^*+\bdelta-\bleta)}\nonumber\\
	&&-\bigg[\ell(\Theta,p)+(\bleta^*-\bleta)\trans\nabla_1(\bleta|\bgamma)+\mfrac{1}{2}(\bleta^*-\bleta)\trans c_1\nabla_2(\bleta|\bgamma)(\bleta^*-\bleta)\bigg],\nonumber\\
	&=& \bdelta\trans\nabla_1(\bleta|\bgamma)+\mfrac{c_1}{2}\bdelta\trans \nabla_2(\bleta|\bgamma)(\bleta^*-\bleta),\nonumber\\
	&=& -\bdelta\trans(\lambda_1{\bf I_0}+\lambda_2{\bf W_0})\bleta^*-\mfrac{c_1}{2}\bdelta\trans\nabla_2(\bleta|\bgamma)(\bleta^*-\bleta)\nonumber.
\end{eqnarray}
The last equality follows from eq. (\ref{eq::sol-eta}), by observing that
\begin{align*}
	\bdelta\trans\nabla_1(\bleta|\bgamma)=-\bdelta\trans(\lambda_1{\bf I_0}+\lambda_2{\bf W_0})\bleta^*-c_1\bdelta\trans\nabla_2(\bleta|\bgamma)(\bleta^*-\bleta).
\end{align*}
%Assuming linear constraint ${\bf B}\trans\bleta={\bf d}$, from eq. (\ref{eq::sol-eta}) we also have ${\bf B}\trans\bleta^*={\bf d}$. 
Substituting $\bdelta=\bleta-\bleta^*$ and noting that $\mathcal{L}_p(\bleta|\bleta)+P(\bleta;\lambda_1,\lambda_2)=F(\bleta|\bgamma,p)$ we have,
\begin{eqnarray}
	F(\bleta|\bgamma,p)-F(\bleta^*|\bgamma,p) &=& \mathcal{L}_p(\bleta|\bleta,\bgamma)-\mathcal{L}_p(\bleta^*|\bleta,\bgamma)+P(\bleta;\lambda_1,\lambda_2)-P(\bleta^*;\lambda_1,\lambda_2),\nonumber\\
	&=&-(\bleta-\bleta^*)\trans(\lambda_1{\bf I_0}+\lambda_2{\bf W_0})\bleta^*-\mfrac{c_1}{2}(\bleta-\bleta^*)\trans\nabla_2(\bleta|\bgamma)(\bleta^*-\bleta)\nonumber\\
	&&+\mfrac{1}{2}\bleta\trans(\lambda_1{\bf I_0}+\lambda_2{\bf W_0})\bleta-\mfrac{1}{2}{\bleta^*}\trans(\lambda_1{\bf I_0}+\lambda_2{\bf W_0})\bleta^*,\nonumber\\
	&=&\mfrac{1}{2}(\bleta^*-\bleta)\trans\big(\lambda_1{\bf I_0}+\lambda_2{\bf W_0}+c_1\nabla_2(\bleta|\bgamma)\big)(\bleta^*-\bleta),\nonumber\\
	&=& \mfrac{\lambda_1}{2}(\bleta^*-\bleta)\trans{\bf I_0}(\bleta^*-\bleta)+\mfrac{1}{2}(\bleta^*-\bleta)\trans\big(\lambda_2{\bf W_0}+c_1\nabla_2(\bleta|\bgamma)\big)(\bleta^*-\bleta),\nonumber\\
	&\geq&\mfrac{\lambda_1}{2}(\bleta^*-\bleta)\trans{\bf I_0}(\bleta^*-\bleta).\nonumber
\end{eqnarray}
 The last inequality follows noting that there exists $c_1$ such that $\lambda_2{\bf W_0}+c_1\nabla_2(\bleta|\bgamma)$ is p.s.d. Under proper choice of constant $c_1$, $\mathcal{L}_p(\bleta^*|\bleta,\bgamma)$ serves as the majorizing function. Existence of $\mathcal{L}_p(\bleta^*|\bleta,\bgamma)$ and $c_1$ is guaranteed by Taylor approximation for multi-variable functions.\\

\noindent {\em Proof of Theorem 2:}\\

\noindent Quantities $\nabla_1(\bleta|\bgamma)$ and $\nabla_2(\bgamma|\bleta)$ necessary for proposed algorithm (\ref{algo::md-1}), are obtained from the negative log-likelihood in eq. (\ref{eq::neg-log-lik-tw-ed}). The partitioned gradient vector,

\begin{subequations}\label{eq::neg-lik-deriv}
	\begin{align}\label{eq::mean-d1}
	\nabla_1(\bleta|\bgamma)=\del{\ell(\Theta,p)}{\bleta}=\begin{cases}
	\del{\ell(\Theta)}{\beta_{k_1}}=-\sum\limits_{i}\sum\limits_{j}x_{ijk_1}D'_{ij}(\bleta)/h_2(z\trans_{ij}\small \bgamma)&\\
	\del{\ell(\Theta)}{\alpha_i}=-\sum\limits_{j}r_{ij}D'_{ij}(\bleta)/h_2(z\trans_{ij}\small \bgamma)& 
	\end{cases},
	\end{align}
	where $D'_{ij}(\cdot)=\Big[k'_{h_1}(\cdot)h'_1(\cdot)\big\{y_{ij}-\kappa'(\cdot)\big\}\Big]$, $(xr)_{ij}\trans\bleta=(x_{ij}|r_{ij})\trans\bleta=(x_{ij}\trans{\small \bbeta}+r_{ij}\trans{\small \balpha})$, $D'_{ij}(\bleta)\equiv D'_{ij}\big((xr)_{ij}\trans\bleta\big)$. Blocks of the Hessian matrix,
	\begin{align}\label{eq::mean-d2}
	\nabla_2(\bleta|\bgamma)=\mfrac{\partial^2\ell(\Theta,p)}{\partial\bleta\partial\bleta\trans}=\begin{cases}	\mfrac{\partial^2\ell(\Theta)}{\partial\beta_{k_1}\partial\beta_{k'_1}}=-\sum\limits_{i}\sum\limits_{j}x_{ijk'_1}D''_{ij}(\bleta)x_{ijk_1}/h_2(z\trans_{ij}\small \bgamma)& \\
	\mfrac{\partial^2\ell(\Theta)}{\partial\alpha_i\partial\alpha_{i'}}=-\sum\limits_{j}r_{i'j}\trans D''_{ij}(\bleta)r_{ij}/h_2(z\trans_{ij}\small \bgamma)& \\
	\mfrac{\partial^2\ell(\Theta)}{\partial\beta_{k_1}\partial\alpha_i}=-\sum\limits_{j}r_{ij}\trans D''_{ij}(\bleta)x_{ijk_1}/h_2(z\trans_{ij}\small \bgamma)&
	\end{cases},
	\end{align}
	where $k_1,k_1'=1,\ldots,k_\beta$, $i,i'=1,\ldots,L$ and
	\begin{align*}
	D''_{ij}(\cdot)=\Big[\big(k''_{h_1}(\cdot)h'^2_1(\cdot)+k'_{h_1}(\cdot)h''_1(\cdot)\big)\big\{y_{ij}-\kappa'(\cdot)\big\}-k'_{h_1}(\cdot)h'_1(\cdot)\kappa''(\cdot)\Big],
	\end{align*}
	and $D''_{ij}(\bleta)\equiv D''_{ij}\big((xr)_{ij}\trans\bleta\big)$. Finally if 
	\begin{align}\label{eq::tw-dev}
	D_{ij}(\cdot) = \Big[y_{ij}k_{h_1}(\cdot)-\kappa\big(k_{h_1}(\cdot)\big)\Big],
	\end{align}
	and $D_{ij}(\bleta)\equiv D_{ij}\big((xr)_{ij}\trans\bleta\big)$, 
	then for $k_2,k'_2=1,\ldots,k_\gamma$ we have the gradient vector,
	\begin{align}\label{eq::disp-d1}
	\nabla_1(\bgamma|\bleta)=\del{\ell(\Theta,p)}{\gamma_{k_2}}=-\sum\limits_{i}\sum\limits_{j}z_{ijk_2}\Bigg\{D_{ij}(\bleta)\Bigg(-\mfrac{h_2'(z\trans_{ij}\small \bgamma)}{h^2_2(z\trans_{ij}\small \bgamma)}\Bigg)+\frac{a'_{ij}(\bgamma)}{a_{ij}(\bgamma)}\Bigg\},
	\end{align}
	where, $a_{ij}(\cdot)=a\big(y_{ij},h_2(\cdot),p\big)$ and $a'_{ij}(\cdot)=a'\big(y_{ij},h_2(\cdot),p\big).h_2'(\cdot)$ with $a_{ij}(\bgamma)\equiv a_{ij}(z_{ij}\trans\bgamma)$, $a'_{ij}(\bgamma)\equiv a'_{ij}(z_{ij}\trans\bgamma)$. The Hessian is given by,
	\begin{align}\label{eq::disp-d2}
	\scalemath{0.9}{\nabla_2(\bgamma|\bleta)=\mfrac{\partial^2\ell(\Theta,p)}{\partial\gamma_{k_2}\partial\gamma_{k'_2}}=-\sum\limits_{i}\sum\limits_{j}z_{ijk_2}\Bigg\{D_{ij}(\bleta)\Bigg(2\mfrac{h_2'^2(z\trans_{ij}\small \bgamma)}{h^3_2(z\trans_{ij}\small \bgamma)}-\mfrac{h_2''(z\trans_{ij}\small \bgamma)}{h^2_2(z\trans_{ij}\small \bgamma)}\Bigg)-\frac{a'^2_{ij}(\bgamma)}{a^2_{ij}(\bgamma)}+\frac{a''_{ij}(\bgamma)}{a_{ij}(\bgamma)}\Bigg\}z_{ijk'_2}},
	\end{align}
\end{subequations}
where, $a''_{ij}(\cdot)=a''\big(y_{ij},h_2(\cdot),p\big).h_2'^2(\cdot)+a'\big(y_{ij},h_2(\cdot),p\big).h_2''(\cdot)$ and $a''_{ij}(\bgamma)\equiv a''_{ij}(z_{ij}\trans\bgamma)$. Naturally, if $\nabla_1(\bleta|\bgamma)$, $\nabla_2(\bleta|\bgamma)$, $\nabla_1(\bgamma|\bleta)$ and $\nabla_2(\bgamma|\bleta)$ in eqs. (\ref{eq::neg-lik-deriv}a--e) exist, we then possess necessary machinery for algorithm (\ref{algo::md-1}) to work.\\

\noindent {\em Proof of Lemma 1}\\

\noindent Following (Dunn and Smyth (2005) \cite{dunn2005series}), we note that,
\begin{gather*}
a(\bgamma)=y^{-1}\sum\limits_{k=1}^{\infty}a_k(y,p)\bigg\{h_2(z\trans\small \bgamma)\bigg\}^{-k(1+\xi)}I(y>0)+1\cdot I(y=0),\\
a'(\bgamma)=-y^{-1}\sum\limits_{k=1}^{\infty}k(1+\xi)a_k(y,p)\bigg\{h_2(z\trans\small \bgamma)\bigg\}^{-k(1+\xi)-1}h'_2(z\trans\small \bgamma)I(y>0)+0\cdot I(y=0),\\
\scalemath{0.9}{a''(\bgamma)=y^{-1}\sum\limits_{k=1}^{\infty}\Bigg[k(1+\xi)a_k(y,p)h_2(z\trans\small \bgamma)^{-k(1+\xi)-1}\bigg\{\big(k(1+\xi)-1\big)\frac{h'_2(z\trans\small \bgamma)^2}{h_2(z\trans\small \bgamma)}-h''_2(z\trans\small \bgamma)\bigg\}\Bigg]I(y>0)+0\cdot I(y=0)},
\end{gather*}
where $a_k(y,p)=\mfrac{y^{k\xi}(p-1)^{-k\xi}}{(2-p)^{k}k!\Gamma(k\xi)}$. Define $t(\bgamma)=\mfrac{y^{\xi}}{(p-1)^{\xi}(2-p)}h_2(z\trans\small\bgamma)^{-(1+\xi)}$. If $T_k(\bgamma)=a_k(y,p)\bigg\{h_2(z\trans\small \bgamma)\bigg\}^{-k(1+\xi)}$ then,
 \begin{align*}
 	\log\big(T_k(\bgamma)\big)=k\log t(\bgamma)-\log\Gamma(k+1)-\log(\Gamma(k\xi)).
 \end{align*}
For large $k$, using Stirling's approximation for Gamma function (see \cite{abramowitz1965handbook}) and setting $k\xi-1\approx k\xi$ we have,
 \begin{align*}
 	\log T_k(\bgamma)\approx k\{\log t(\bgamma)+(1+\xi)-\xi\log\xi-(1+\xi)\log k\}-\log k -\mfrac{1}{2}\log\xi-\log 2\pi.
 \end{align*}
Differentiating with respect to $k$ and ignoring $\mfrac{1}{k}$ (since  $k$ is large) we have,
\begin{align*}
	\mfrac{\partial \log T_k(\bgamma)}{\partial k} \approx \log t(\bgamma)-\xi\log\xi k-\log k.
\end{align*}
Finally solving $\mfrac{\partial T_k(\bgamma)}{\partial k}=0$, yields $k_{\max}=\mfrac{y^{2-p}}{2-p}h_2(z\trans\small\bgamma)^{-1}$.

 For $a'(\bgamma)$, if we denote the summands by $T'_k(\bgamma)=k(1+\xi)a_k(y,p)\bigg\{h_2(z\trans\small \bgamma)\bigg\}^{-k(1+\xi)}\mfrac{h'_2(z\trans\small \bgamma)}{h_2(z\trans\small \bgamma)}$ we have,
\begin{align*}
	\log T'_k(\bgamma)=\log k+\log(1+\xi)+k\log t(\bgamma)-\log\Gamma(k+1)-\log\Gamma(k\xi)-\log\mfrac{h'_2(z\trans\small \bgamma)}{h_2(z\trans\small \bgamma)}.
\end{align*}
Using Stirling's approximation followed by differentiating with respect to $k$ we have $k_{\max}=\mfrac{y^{2-p}}{2-p}h_2(z\trans\small\bgamma)^{-1}$. Similar calculations follow for $a''(\bgamma)$ completing the proof. 

\newpage

\section*{Appendix B: Tables}

\vspace*{2cm}

\begin{table}[H]
\centering
\caption{Comparative results over 10 replications for the un-penalized ($\Theta_0$), ridge ($\Theta_r$) and estimates from algorithm \ref{algo::md-1} ($\Theta$) using metrics described in section \ref{sec::sim}, are shown for a \emph{block spatial pattern} under varying proportions of zeros (SNRs) and sample sizes in synthetic response. Respective standard deviations are shown in brackets below the value of an error metric. A dominant result in favor of the proposed estimates is shown in bold.}\label{tab::sim-bl}
\resizebox{\linewidth}{!}{
	\begin{tabular}{|c|c|@{\extracolsep{8pt}}cccccccccccc|@{}}
	 \hline
	 \hline
	 \multirow{7}{*}{Sample Size} & \multirow{7}{2.7cm}{\centering Prop. of zeros (SNR)} & \multicolumn{9}{c}{\multirow{3}{*}{SSE}}  & \multicolumn{3}{c|}{\multirow{3}{*}{Deviance Ratio (Validation)}} \\ 
	 &&&&&&&&&&&&&\\
	 &&&&&&&&&&&&&\\
	 \cline{3-11}\cline{12-14}
	 &&\multicolumn{3}{c}{\multirow{2}{*}{Mean}}&\multicolumn{3}{c}{\multirow{2}{*}{Spatial Effect}}&\multicolumn{3}{c}{\multirow{2}{*}{Dispersion}}&\multirow{4}{*}{$\Theta_0$}&\multirow{4}{*}{$\Theta_r$}&\multirow{4}{*}{$\Theta$}\\
	 &&&&&&&&&&&&&\\
	 \cline{3-5}\cline{6-8}\cline{9-11}
	 &&\multirow{2}{*}{$\bbeta_0$}&\multirow{2}{*}{$\bbeta_r$}&\multirow{2}{*}{$\bbeta$}&\multirow{2}{*}{$\balpha_0$}&\multirow{2}{*}{$\balpha_r$}&\multirow{2}{*}{$\balpha$}&\multirow{2}{*}{$\bgamma_0$}&\multirow{2}{*}{$\bgamma_r$}&\multirow{2}{*}{$\bgamma$}&&&\\
	 &&&&&&&&&&&&&\\
	 \hline
	 &&&&&&&&&&&&&\\
	\multirow{12}{*}{10000}& \multirow{2}{*}{0.15} & 0.008 & 0.009 & 0.009 & 25.669 & 19.270 & \bf 11.840 & 0.110 & 0.108 & 0.109 & 1.046 & 0.958 & \bf 1.041 \\ 
	&  & (0.006) & (0.007) & (0.008) & (5.366) & (2.573) & (1.197) & (0.044) & (0.044) & (0.046) & (0.005) & (0.003) & (0.004) \\ 
	&&&&&&&&&&&&&\\
	& \multirow{2}{*}{0.30} & 0.111 & 0.094 & 0.095 & 42.894 & 29.519 & \bf 18.843 & 0.025 & 0.026 & 0.027 & 1.054 & 0.964 & \bf 1.041 \\ 
	& & (0.043) & (0.038) & (0.043) & (9.800) & (3.265) & (1.405) & (0.003) & (0.004) & (0.003) & (0.006) & (0.003) & (0.005) \\
	&&&&&&&&&&&&&\\
	& \multirow{2}{*}{0.60} & 0.677 & 0.054 & \bf 0.048 & 2742.248 & 94.28 & \bf 43.718 & 0.129 & 0.097 & 0.100 & 1.074 & 0.964 & \bf 1.044 \\ 
	& & (0.227) & (0.034) & (0.046) & (547.623) & (4.220) & (2.882) & (0.057) & (0.042) & (0.057) & (0.010) & (0.011) & (0.007) \\ 
	&&&&&&&&&&&&&\\
	&\multirow{2}{*}{0.80} & 7.374 & 0.219 & \bf 0.150 & 13231.464 & 233.888 & \bf 70.388 & 0.852 & 0.742 & \bf 0.741 & 1.118 & 0.975 & \bf 1.053 \\ 
	& & (1.381) & (0.158) & (0.120) & (1159.38) & (59.698) & (3.799) & (0.105) & (0.104) & (0.094) & (0.013) & (0.003) & (0.009) \\ 
	&&&&&&&&&&&&&\\
	\hline
	&&&&&&&&&&&&&\\
	\multirow{12}{*}{20000} & \multirow{2}{*}{0.15} & 0.004 & 0.005 & 0.005 & 7.189 & 7.074 & \bf 6.015 & 0.131 & 0.132 & 0.134 & 1.021 & 0.979 & \bf 1.020 \\ 
	& & (0.003) & (0.002) & (0.002) & (0.900) & (0.589) & (0.562) & (0.015) & (0.015) & (0.016) & (0.004) & (0.002) & (0.003) \\ 
	&&&&&&&&&&&&&\\
	&\multirow{2}{*}{0.30} & 0.006 & 0.010 & 0.008 & 18.447 & 16.136 & \bf 11.998 & 0.049 & 0.049 & 0.050 & 1.022 & 0.978 & \bf 1.018 \\ 
	& & (0.006) & (0.008) & (0.007) & (2.235) & (0.884) & (0.688) & (0.018) & (0.018) & (0.018) & (0.005) & (0.002) & (0.004) \\ 
	&&&&&&&&&&&&&\\
	& \multirow{2}{*}{0.60} & 0.112 & 0.028 & 0.029 & 621.235 & 54.711 & \bf 29.675 & 0.081 & 0.076 & \bf 0.071 & 1.032 & 0.975 & \bf 1.024 \\ 
	& & (0.079) & (0.024) & (0.031) & (338.214) & (5.594) & (2.397) & (0.019) & (0.018) & (0.015) & (0.007) & (0.002) & (0.005) \\ 
	&&&&&&&&&&&&&\\
	& \multirow{2}{*}{0.80} & 1.456 & 0.080 & \bf 0.070 & 5778.077 & 121.449 & \bf 55.990 & 0.633 & 0.612 & \bf 0.578 & 1.049 & 0.973 & \bf 1.025 \\ 
	& & (0.323) & (0.109) & (0.091) & (983.308) & (4.747) & (4.814) & (0.066) & (0.065) & (0.053) & (0.009) & (0.006) & (0.006) \\ 
	&&&&&&&&&&&&&\\
	\hline
	&&&&&&&&&&&&&\\
	\multirow{12}{*}{30000} & \multirow{2}{*}{0.15} & 0.007 & 0.007 & 0.007 & 5.232 & 5.015 & \bf 4.387 & 0.164 & 0.162 & 0.163 & 1.013 & 0.984 & \bf 1.012 \\ 
	& & (0.006) & (0.006) & (0.006) & (0.656) & (0.652) & (0.397) & (0.020) & (0.017) & (0.016) & (0.003) & (0.001) & (0.003) \\ 
	&&&&&&&&&&&&&\\
	& \multirow{2}{*}{0.30} & 0.005 & 0.006 & \bf 0.005 & 14.015 & 11.757 & \bf 8.753 & 0.041 & 0.042 & 0.042 & 1.015 & 0.986 & \bf 1.014 \\ 
	& & (0.005) & (0.005) & (0.005) & (1.925) & (1.326) & (0.859) & (0.009) & (0.010) & (0.009) & (0.002) & (0.001) & (0.002) \\ 
	&&&&&&&&&&&&&\\
	& \multirow{2}{*}{0.60} & 0.026 & 0.014 & \bf 0.011 & 218.063 & 33.548 & \bf 24.691 & 0.077 & 0.075 & \bf 0.073 & 1.019 & 0.982 & \bf 1.015 \\ 
	& & (0.018) & (0.007) & (0.006) & (212.702) & (7.025) & (2.706) & (0.033) & (0.032) & (0.032) & (0.003) & (0.001) & (0.003) \\ 
	&&&&&&&&&&&&&\\
	& \multirow{2}{*}{0.80} & 0.293 & 0.033 & 0.039 & 2333.53 & 77.805 & \bf 37.706 & 0.544 & 0.533 & \bf 0.505 & 1.033 & 0.985 & \bf 1.022 \\ 
	& & (0.149) & (0.021) & (0.026) & (559.65) & (4.163) & (2.179) & (0.032) & (0.03) & (0.024) & (0.006) & (0.002) & (0.005) \\
	&&&&&&&&&&&&&\\
	\hline
	&&&&&&&&&&&&&\\ 
	\multirow{12}{*}{50000} & \multirow{2}{*}{0.15} & 0.001 & 0.002 & 0.002 & 3.173 & 3.112 & \bf 2.977 & 0.145 & 0.145 & \bf 0.145 & 1.007 & 0.990 & \bf 1.007 \\ 
	& & (0.001) & (0.001) & (0.001) & (0.262) & (0.237) & (0.179) & (0.008) & (0.008) & (0.008) & (0.001) & (0.000) & (0.001) \\ 
	&&&&&&&&&&&&&\\ 
	& \multirow{2}{*}{0.30} & 0.003 & 0.005 & 0.004 & 5.887 & 5.647 & \bf 5.059 & 0.051 & 0.052 & \bf 0.051 & 1.009 & 0.991 & \bf 1.009 \\ 
	& & (0.002) & (0.003) & (0.002) & (0.824) & (0.637) & (0.392) & (0.006) & (0.006) & (0.006) & (0.001) & (0.001) & (0.001) \\ 
	&&&&&&&&&&&&&\\ 
	& \multirow{2}{*}{0.60} & 0.008 & 0.009 & \bf 0.007 & 26.591 & 21.627 & \bf 16.823 & 0.046 & 0.045 & \bf 0.044 & 1.011 & 0.987 & \bf 1.009 \\ 
	& & (0.008) & (0.006) & (0.007) & (2.989) & (2.137) & (1.543) & (0.006) & (0.006) & (0.006) & (0.002) & (0.001) & (0.002) \\ 
	&&&&&&&&&&&&&\\ 
	& \multirow{2}{*}{0.80} & 0.058 & 0.030 & 0.032 & 744.269 & 49.523 & \bf 28.606 & 0.519 & 0.518 & \bf 0.514 & 1.021 & 0.987 & \bf 1.014 \\ 
	& & (0.023) & (0.019) & (0.019) & (355.157) & (6.689) & (2.353) & (0.022) & (0.021) & (0.021) & (0.004) & (0.001) & (0.003) \\ 
	&&&&&&&&&&&&&\\ 
	\hline
	\hline
	\end{tabular}
}
\end{table}

\newpage

\vspace*{3cm}

\begin{table}[H]
	\centering
	\caption{Comparative results over 10 replications for the un-penalized ($\Theta_0$), ridge ($\Theta_r$) and estimates from algorithm \ref{algo::md-1} ($\Theta$) using metrics described in section \ref{sec::sim}, are shown for a \emph{smooth spatial pattern} under varying proportions of zeros (SNRs) and sample sizes in synthetic response. Respective standard deviations are shown in brackets below the value of an error metric. A dominant result in favor of the proposed estimates is shown in bold.}\label{tab::sim-sm}
	\resizebox{\linewidth}{!}{
		\begin{tabular}{|c|c|@{\extracolsep{8pt}}cccccccccccc|@{}}
			\hline
			\hline
			\multirow{7}{*}{Sample Size} & \multirow{7}{2.7cm}{\centering Prop. of zeros (SNR)} & \multicolumn{9}{c}{\multirow{3}{*}{SSE}}  & \multicolumn{3}{c|}{\multirow{3}{*}{Deviance Ratio (Validation)}} \\ 
			&&&&&&&&&&&&&\\
			&&&&&&&&&&&&&\\
			\cline{3-11}\cline{12-14}
			&&\multicolumn{3}{c}{\multirow{2}{*}{Mean}}&\multicolumn{3}{c}{\multirow{2}{*}{Spatial Effect}}&\multicolumn{3}{c}{\multirow{2}{*}{Dispersion}}&\multirow{4}{*}{$\Theta_0$}&\multirow{4}{*}{$\Theta_r$}&\multirow{4}{*}{$\Theta$}\\
			&&&&&&&&&&&&&\\
			\cline{3-5}\cline{6-8}\cline{9-11}
			&&\multirow{2}{*}{$\bbeta_0$}&\multirow{2}{*}{$\bbeta_r$}&\multirow{2}{*}{$\bbeta$}&\multirow{2}{*}{$\balpha_0$}&\multirow{2}{*}{$\balpha_r$}&\multirow{2}{*}{$\balpha$}&\multirow{2}{*}{$\bgamma_0$}&\multirow{2}{*}{$\bgamma_r$}&\multirow{2}{*}{$\bgamma$}&&&\\
			&&&&&&&&&&&&&\\
			\hline
			&&&&&&&&&&&&&\\
			\multirow{12}{*}{10000}& \multirow{2}{*}{0.15} & 0.017 & 0.023 & 0.021 & 16.154 & 14.406 & \bf 7.458 & 0.107 & 0.112 & 0.119 & 1.043 & 0.968 & \bf1.031 \\ 
			&  & (0.014) & (0.017) & (0.014) & (2.837) & (1.766) & (0.475) & (0.017) & (0.016) & (0.013) & (0.009) & (0.003) & (0.007) \\ 
			&&&&&&&&&&&&&\\
			& \multirow{2}{*}{0.30} & 0.023 & 0.036 & 0.029 & 49.547 & 27.846 & \bf 11.44 & 0.044 & 0.046 & 0.048 & 1.050 & 0.968 & \bf 1.030 \\ 
			& & (0.015) & (0.023) & (0.018) & (16.689) & (2.989) & (1.124) & (0.018) & (0.016) & (0.018) & (0.009) & (0.003) & (0.006) \\
			&&&&&&&&&&&&&\\
			& \multirow{2}{*}{0.60} & 0.548 & 0.097 & \bf 0.096 & 2073.488 & 76.937 & \bf 21.447 & 0.105 & 0.091 & \bf 0.074 & 1.059 & 0.963 & \bf 1.021 \\ 
			& & (0.196) & (0.053) & (0.044) & (756.98) & (2.548) & (1.846) & (0.040) & (0.035) & (0.028) & (0.007) & (0.002) & (0.004) \\ 
			&&&&&&&&&&&&&\\
			&\multirow{2}{*}{0.80} & 3.672 & 0.154 & \bf 0.095 & 8806.711 & 212.408 & \bf 35.205 & 0.786 & 0.632 & 0.652 & 1.094 & 0.953 & \bf 1.029 \\ 
			& & (0.866) & (0.102) & (0.077) & (1117.467) & (40.505) & (3.675) & (0.135) & (0.12) & (0.086) & (0.014) & (0.004) & (0.009) \\ 
			&&&&&&&&&&&&&\\
			\hline
			&&&&&&&&&&&&&\\
			\multirow{12}{*}{20000} & \multirow{2}{*}{0.15} & 0.008 & 0.010 & 0.010 & 8.327 & 7.422 & \bf 4.403 & 0.179 & 0.180 & 0.180 & 1.020 & 0.981 & \bf 1.015 \\ 
			& & (0.006) & (0.006) & (0.007) & (1.097) & (0.93) & (0.413) & (0.023) & (0.023) & (0.022) & (0.003) & (0.002) & (0.004) \\ 
			&&&&&&&&&&&&&\\
			&\multirow{2}{*}{0.30} & 0.025 & 0.032 & 0.029 & 17.831 & 13.990 & \bf 7.643 & 0.071 & 0.071 & 0.072 & 1.020 & 0.981 & \bf 1.014 \\ 
			& & (0.021) & (0.022) & (0.021) & (1.969) & (1.091) & (0.817) & (0.023) & (0.022) & (0.022) & (0.004) & (0.001) & (0.004) \\ 
			&&&&&&&&&&&&&\\
			& \multirow{2}{*}{0.60} & 0.058 & 0.085 & 0.073 & 123.484 & 38.731 & \bf 15.069 & 0.046 & 0.044 & \bf 0.040 & 1.027 & 0.979 & \bf 1.014 \\ 
			& & (0.037) & (0.048) & (0.043) & (132.932) & (2.881) & (1.18) & (0.018) & (0.017) & (0.013) & (0.004) & (0.002) & (0.003) \\ 
			&&&&&&&&&&&&&\\
			& \multirow{2}{*}{0.80} & 0.770 & 0.228 & \bf 0.202 & 2577.69 & 95.043 & \bf 27.924 & 0.642 & 0.623 & \bf 0.574 & 1.047 & 0.973 & \bf 1.017 \\ 
			& & (0.332) & (0.182) & (0.161) & (931.672) & (14.554) & (3.341) & (0.115) & (0.108) & (0.094) & (0.012) & (0.01) & (0.008) \\ 
			&&&&&&&&&&&&&\\
			\hline
			&&&&&&&&&&&&&\\
			\multirow{12}{*}{30000} & \multirow{2}{*}{0.15} & 0.003 & 0.003 & 0.003 & 4.689 & 4.603 & \bf 3.154 & 0.130 & 0.131 & 0.132 & 1.014 & 0.987 & \bf 1.012 \\ 
			& & (0.002) & (0.002) & (0.002) & (0.625) & (0.560) & (0.271) & (0.013) & (0.013) & (0.013) & (0.003) & (0.001) & (0.002) \\ 
			&&&&&&&&&&&&&\\
			& \multirow{2}{*}{0.30} & 0.005 & 0.007 & 0.006 & 9.156 & 9.145 & \bf 5.509 & 0.055 & 0.055 & \bf 0.053 & 1.013 & 0.986 & \bf 1.010 \\ 
			& & (0.006) & (0.005) & (0.006) & (0.596) & (0.528) & (0.243) & (0.015) & (0.015) & (0.015) & (0.002) & (0.001) & (0.002) \\ 
			&&&&&&&&&&&&&\\
			& \multirow{2}{*}{0.60} & 0.018 & 0.029 & 0.023 & 178.556 & 28.990 & \bf 13.413 & 0.033 & 0.032 & 0.033 & 1.020 & 0.983 & \bf 1.012 \\ 
			& & (0.010) & (0.016) & (0.014) & (159.821) & (1.617) & (1.251) & (0.007) & (0.007) & (0.008) & (0.003) & (0.001) & (0.002) \\ 
			&&&&&&&&&&&&&\\
			& \multirow{2}{*}{0.80} & 0.096 & 0.038 & \bf 0.028 & 853.972 & 73.535 & \bf 21.599 & 0.489 & 0.481 & \bf 0.470 & 1.035 & 0.985 & \bf 1.013 \\ 
			& & (0.059) & (0.018) & (0.011) & (625.431) & (12.347) & (2.046) & (0.046) & (0.046) & (0.041) & (0.005) & (0.007) & (0.004) \\
			&&&&&&&&&&&&&\\
			\hline
			&&&&&&&&&&&&&\\ 
			\multirow{12}{*}{50000} & \multirow{2}{*}{0.15} & 0.002 & 0.003 & 0.003 & 3.063 & 2.926 & \bf 2.422 & 0.153 & 0.153 & 0.154 & 1.007 & 0.989 & \bf 1.006 \\ 
			& & (0.001) & (0.001) & (0.001) & (0.186) & (0.158) & (0.115) & (0.014) & (0.014) & (0.014) & (0.001) & (0.001) & (0.001) \\ 
			&&&&&&&&&&&&&\\ 
			& \multirow{2}{*}{0.30} & 0.004 & 0.006 & 0.005 & 6.020 & 5.615 & \bf 4.172 & 0.042 & 0.042 & 0.042 & 1.009 & 0.991 & \bf 1.007 \\ 
			& & (0.004) & (0.005) & (0.004) & (0.617) & (0.603) & (0.343) & (0.009) & (0.009) & (0.009) & (0.002) & (0.001) & (0.002) \\ 
			&&&&&&&&&&&&&\\ 
			& \multirow{2}{*}{0.60} & 0.007 & 0.012 & 0.011 & 23.661 & 18.238 & \bf 9.705 & 0.035 & 0.034 & \bf 0.033 & 1.011 & 0.990 & \bf 1.007 \\ 
			& & (0.005) & (0.005) & (0.005) & (2.16) & (1.292) & (0.888) & (0.007) & (0.007) & (0.006) & (0.002) & (0.003) & (0.002) \\ 
			&&&&&&&&&&&&&\\ 
			& \multirow{2}{*}{0.80} & 0.074 & 0.047 & \bf 0.042 & 256.321 & 41.535 & \bf 17.700 & 0.482 & 0.478 & \bf 0.457 & 1.018 & 0.990 & \bf 1.009 \\ 
			& & (0.054) & (0.040) & (0.033) & (205.412) & (3.168) & (1.226) & (0.022) & (0.021) & (0.019) & (0.002) & (0.001) & (0.001) \\ 
			&&&&&&&&&&&&&\\ 
			\hline
			\hline
		\end{tabular}
	}
\end{table}

\newpage

\vspace*{3cm}

\begin{table}[H]
	\centering
	\caption{Comparative results over 10 replications for the un-penalized ($\Theta_0$), ridge ($\Theta_r$) and estimates from algorithm \ref{algo::md-1} ($\Theta$) using metrics described in section \ref{sec::sim}, are shown for a \emph{hotspot spatial pattern} under varying proportions of zeros (SNRs) and sample sizes in synthetic response. Respective standard deviations are shown in brackets below the value of an error metric. A dominant result in favor of the proposed estimates is shown in bold.}\label{tab::sim-hs}
	\resizebox{\linewidth}{!}{
		\begin{tabular}{|c|c|@{\extracolsep{8pt}}cccccccccccc|@{}}
			\hline
			\hline
			\multirow{7}{*}{Sample Size} & \multirow{7}{2.7cm}{\centering Prop. of zeros (SNR)} & \multicolumn{9}{c}{\multirow{3}{*}{SSE}}  & \multicolumn{3}{c|}{\multirow{3}{*}{Deviance Ratio (Validation)}} \\ 
			&&&&&&&&&&&&&\\
			&&&&&&&&&&&&&\\
			\cline{3-11}\cline{12-14}
			&&\multicolumn{3}{c}{\multirow{2}{*}{Mean}}&\multicolumn{3}{c}{\multirow{2}{*}{Spatial Effect}}&\multicolumn{3}{c}{\multirow{2}{*}{Dispersion}}&\multirow{4}{*}{$\Theta_0$}&\multirow{4}{*}{$\Theta_r$}&\multirow{4}{*}{$\Theta$}\\
			&&&&&&&&&&&&&\\
			\cline{3-5}\cline{6-8}\cline{9-11}
			&&\multirow{2}{*}{$\bbeta_0$}&\multirow{2}{*}{$\bbeta_r$}&\multirow{2}{*}{$\bbeta$}&\multirow{2}{*}{$\balpha_0$}&\multirow{2}{*}{$\balpha_r$}&\multirow{2}{*}{$\balpha$}&\multirow{2}{*}{$\bgamma_0$}&\multirow{2}{*}{$\bgamma_r$}&\multirow{2}{*}{$\bgamma$}&&&\\
			&&&&&&&&&&&&&\\
			\hline
			&&&&&&&&&&&&&\\
			\multirow{12}{*}{10000}& \multirow{2}{*}{0.15} & 0.014 & 0.016 & \bf 0.013 & 14.556 & 10.915 & \bf 2.697 & 0.048 & 0.049 & 0.054 & 1.035 & 0.986 & \bf 1.009 \\ 
			&  & (0.008) & (0.01) & (0.01) & (1.438) & (0.879) & (0.412) & (0.011) & (0.007) & (0.007) & (0.007) & (0.005) & (0.003) \\ 
			&&&&&&&&&&&&&\\
			& \multirow{2}{*}{0.30} & 0.021 & 0.035 & 0.026 & 25.794 & 16.255 & \bf 4.162 & 0.030 & 0.029 & 0.032 & 1.034 & 0.977 & \bf 1.009 \\ 
			& & (0.017) & (0.024) & (0.019) & (1.809) & (0.868) & (0.188) & (0.01) & (0.006) & (0.008) & (0.005) & (0.002) & (0.002) \\
			&&&&&&&&&&&&&\\
			& \multirow{2}{*}{0.60} & 0.179 & 0.109 & 0.124 & 67.996 & 26.005 & \bf 5.340 & 0.044 & 0.027 & \bf 0.025 & 1.034 & 0.982 & \bf 1.008 \\ 
			& & (0.087) & (0.065) & (0.076) & (7.815) & (2.077) & (0.668) & (0.015) & (0.01) & (0.011) & (0.008) & (0.007) & (0.003) \\ 
			&&&&&&&&&&&&&\\
			&\multirow{2}{*}{0.80} & 0.368 & 0.279 & \bf 0.254 & 1584.609 & 43.621 & \bf 10.421 & 0.446 & 0.346 & \bf 0.341 & 1.036 & 0.981 & \bf 1.007 \\ 
			& & (0.200) & (0.245) & (0.221) & (455.576) & (2.588) & (1.550) & (0.033) & (0.031) & (0.025) & (0.011) & (0.01) & (0.006) \\ 
			&&&&&&&&&&&&&\\
			\hline
			&&&&&&&&&&&&&\\
			\multirow{12}{*}{20000} & \multirow{2}{*}{0.15} & 0.005 & 0.005 & 0.006 & 6.97 & 5.87 & \bf 1.551 & 0.058 & 0.058 & 0.059 & 1.019 & 0.991 & \bf 1.006 \\ 
			& & (0.003) & (0.002) & (0.003) & (0.423) & (0.449) & (0.136) & (0.006) & (0.006) & (0.005) & (0.003) & (0.001) & (0.001) \\ 
			&&&&&&&&&&&&&\\
			&\multirow{2}{*}{0.30} & 0.008 & 0.009 & \bf 0.007 & 13.784 & 10.787 & \bf 3.225 & 0.035 & 0.037 & 0.038 & 1.015 & 0.986 & \bf 1.004 \\ 
			& & (0.005) & (0.005) & (0.004) & (0.822) & (0.775) & (0.702) & (0.01) & (0.008) & (0.007) & (0.004) & (0.004) & (0.002) \\ 
			&&&&&&&&&&&&&\\
			& \multirow{2}{*}{0.60} & 0.035 & 0.032 & 0.033 & 34.634 & 19.129 & \bf 4.545 & 0.023 & 0.020 & \bf 0.019 & 1.017 & 0.987 & \bf 1.004 \\ 
			& & (0.038) & (0.033) & (0.031) & (2.514) & (1.245) & (1.374) & (0.007) & (0.006) & (0.005) & (0.005) & (0.006) & (0.003) \\ 
			&&&&&&&&&&&&&\\
			& \multirow{2}{*}{0.80} & 0.065 & 0.077 & 0.075 & 81.654 & 32.773 & \bf 6.018 & 0.422 & 0.390 & \bf 0.364 & 1.025 & 0.987 & \bf 1.003 \\ 
			& & (0.038) & (0.060) & (0.052) & (8.731) & (1.857) & (0.801) & (0.037) & (0.033) & (0.031) & (0.006) & (0.004) & (0.003) \\ 
			&&&&&&&&&&&&&\\
			\hline
			&&&&&&&&&&&&&\\
			\multirow{12}{*}{30000} & \multirow{2}{*}{0.15} & 0.003 & 0.003 & 0.003 & 4.475 & 4.122 & \bf 1.234 & 0.093 & 0.096 & 0.098 & 1.012 & 0.993 & \bf 1.004 \\ 
			& & (0.002) & (0.001) & (0.002) & (0.406) & (0.282) & (0.093) & (0.017) & (0.017) & (0.018) & (0.002) & (0.001) & (0.001) \\ 
			&&&&&&&&&&&&&\\
			& \multirow{2}{*}{0.30} & 0.011 & 0.011 & \bf 0.011 & 7.955 & 6.610 & \bf 1.704 & 0.043 & 0.042 & 0.044 & 1.011 & 0.993 & \bf 1.003 \\ 
			& & (0.008) & (0.007) & (0.007) & (0.617) & (0.502) & (0.141) & (0.007) & (0.007) & (0.006) & (0.002) & (0.001) & (0.001)  \\ 
			&&&&&&&&&&&&&\\
			& \multirow{2}{*}{0.60} & 0.024 & 0.025 & 0.030 & 18.576 & 13.558 & \bf 3.146 & 0.015 & 0.014 & \bf 0.013 & 1.015 & 0.994 & \bf 1.003 \\ 
			& & (0.018) & (0.019) & (0.022) & (1.105) & (0.658) & (0.351) & (0.002) & (0.001) & (0.002) & (0.001) & (0.001) & (0.001) \\ 
			&&&&&&&&&&&&&\\
			& \multirow{2}{*}{0.80} & 0.136 & 0.107 & \bf 0.096 & 45.358 & 23.655 & \bf 3.475 & 0.465 & 0.431 & \bf 0.419 & 1.028 & 0.992 & \bf 1.006 \\ 
			& & (0.088) & (0.079) & (0.072) & (4.233) & (1.546) & (0.364) & (0.052) & (0.047) & (0.041) & (0.004) & (0.001) & (0.001) \\
			&&&&&&&&&&&&&\\
			\hline
			&&&&&&&&&&&&&\\ 
			\multirow{12}{*}{50000} & \multirow{2}{*}{0.15} & 0.003 & 0.003 & \bf 0.003 & 2.906 & 2.873 & \bf 1.129 & 0.078 & 0.078 & \bf 0.078 & 1.006 & 0.994 & \bf 1.002 \\ 
			& & (0.003) & (0.003) & (0.003) & (0.185) & (0.165) & (0.078) & (0.011) & (0.011) & (0.010) & (0.001) & (0.000) & (0.001) \\ 
			&&&&&&&&&&&&&\\ 
			& \multirow{2}{*}{0.30} & 0.004 & 0.004 & \bf 0.004 & 4.712 & 4.076 & \bf 1.367 & 0.038 & 0.038 & 0.039 & 1.007 & 0.994 & \bf 1.002 \\ 
			& & (0.003) & (0.003) & (0.002) & (0.292) & (0.252) & (0.134) & (0.003) & (0.003) & (0.003) & (0.001) & (0) & (0.001) \\ 
			&&&&&&&&&&&&&\\ 
			& \multirow{2}{*}{0.60} & 0.009 & 0.010 & 0.010 & 12.981 & 9.836 & \bf 2.16 & 0.020 & 0.019 & \bf 0.018 & 1.008 & 0.994 & \bf 1.003 \\ 
			& & (0.008) & (0.009) & (0.01) & (0.802) & (0.800) & (0.239) & (0.005) & (0.004) & (0.004) & (0.002) & (0.001) & (0.001) \\ 
			&&&&&&&&&&&&&\\ 
			& \multirow{2}{*}{0.80} & 0.032 & 0.033 & \bf 0.029 & 27.714 & 18.677 & \bf 3.589 & 0.395 & 0.392 & \bf 0.378 & 1.013 & 0.992 & \bf 1.003 \\ 
			& & (0.021) & (0.020) & (0.017) & (2.563) & (1.666) & (0.423) & (0.028) & (0.027) & (0.026) & (0.003) & (0.001) & (0.001) \\ 
			&&&&&&&&&&&&&\\ 
			\hline
			\hline
		\end{tabular}
	}
\end{table}

\newpage

\vspace*{3cm}

\begin{table}[H]
	\centering
	\caption{Comparative results over 10 replications for the un-penalized ($\Theta_0$), ridge ($\Theta_r$) and estimates from algorithm \ref{algo::md-1} ($\Theta$) using metrics described in section \ref{sec::sim}, are shown for a \emph{structured spatial pattern} under varying proportions of zeros (SNRs) and sample sizes in synthetic response. Respective standard deviations are shown in brackets below the value of an error metric. A dominant result in favor of the proposed estimates is shown in bold.}\label{tab::sim-str}
	\resizebox{\linewidth}{!}{
		\begin{tabular}{|c|c|@{\extracolsep{8pt}}cccccccccccc|@{}}
			\hline
			\hline
			\multirow{7}{*}{Sample Size} & \multirow{7}{2.7cm}{\centering Prop. of zeros (SNR)} & \multicolumn{9}{c}{\multirow{3}{*}{SSE}}  & \multicolumn{3}{c|}{\multirow{3}{*}{Deviance Ratio (Validation)}} \\ 
			&&&&&&&&&&&&&\\
			&&&&&&&&&&&&&\\
			\cline{3-11}\cline{12-14}
			&&\multicolumn{3}{c}{\multirow{2}{*}{Mean}}&\multicolumn{3}{c}{\multirow{2}{*}{Spatial Effect}}&\multicolumn{3}{c}{\multirow{2}{*}{Dispersion}}&\multirow{4}{*}{$\Theta_0$}&\multirow{4}{*}{$\Theta_r$}&\multirow{4}{*}{$\Theta$}\\
			&&&&&&&&&&&&&\\
			\cline{3-5}\cline{6-8}\cline{9-11}
			&&\multirow{2}{*}{$\bbeta_0$}&\multirow{2}{*}{$\bbeta_r$}&\multirow{2}{*}{$\bbeta$}&\multirow{2}{*}{$\balpha_0$}&\multirow{2}{*}{$\balpha_r$}&\multirow{2}{*}{$\balpha$}&\multirow{2}{*}{$\bgamma_0$}&\multirow{2}{*}{$\bgamma_r$}&\multirow{2}{*}{$\bgamma$}&&&\\
			&&&&&&&&&&&&&\\
			\hline
			&&&&&&&&&&&&&\\
			\multirow{12}{*}{10000}& \multirow{2}{*}{0.15} & 0.023 & 0.029 & 0.033 & 17.411 & 15.883 & \bf 9.346 & 0.081 & 0.081 & 0.082 & 1.040 & 0.981 & \bf 1.020 \\ 
			&  & (0.010) & (0.012) & (0.013) & (1.971) & (1.554) & (0.766) & (0.008) & (0.008) & (0.008) & (0.008) & (0.003) & (0.006) \\ 
			&&&&&&&&&&&&&\\
			& \multirow{2}{*}{0.30} & 0.063 & 0.072 & 0.075 & 28.845 & 24.426 & \bf 13.098 & 0.042 & 0.038 & 0.046 & 1.041 & 0.977 & \bf 1.018 \\ 
			& & (0.051) & (0.054) & (0.056) & (2.899) & (1.798) & (1.884) & (0.032) & (0.027) & (0.032) & (0.009) & (0.009) & (0.006) \\
			&&&&&&&&&&&&&\\
			& \multirow{2}{*}{0.60} & 0.066 & 0.043 & 0.049 & 483.642 & 58.702 & \bf 17.728 & 0.113 & 0.103 & 0.114 & 1.062 & 0.965 & \bf 1.023 \\ 
			& & (0.038) & (0.026) & (0.028) & (428.887) & (5.799) & (2.784) & (0.067) & (0.058) & (0.068) & (0.016) & (0.009) & (0.006) \\ 
			&&&&&&&&&&&&&\\
			&\multirow{2}{*}{0.80} & 2.315 & 0.278 & \bf 0.179 & 7259.813 & 124.73 & \bf 30.908 & 0.628 & 0.502 & 0.544 & 1.084 & 0.965 & \bf 1.025 \\ 
			& & (0.797) & (0.226) & (0.146) & (1441.228) & (7.997) & (3.920) & (0.104) & (0.078) & (0.096) & (0.023) & (0.016) & (0.010) \\ 
			&&&&&&&&&&&&&\\
			\hline
			&&&&&&&&&&&&&\\
			\multirow{12}{*}{20000} & \multirow{2}{*}{0.15} & 0.007 & 0.009 & 0.010 & 7.811 & 7.480 & \bf 5.300 & 0.082 & 0.082 & \bf 0.082 & 1.019 & 0.987 & \bf 1.011 \\ 
			& & (0.005) & (0.006) & (0.005) & (0.948) & (0.723) & (0.526) & (0.007) & (0.007) & (0.007) & (0.004) & (0.001) & (0.002) \\ 
			&&&&&&&&&&&&&\\
			&\multirow{2}{*}{0.30} & 0.004 & 0.005 & 0.007 & 15.458 & 13.580 & \bf 7.715 & 0.048 & 0.047 & 0.048 & 1.023 & 0.986 & \bf 1.011 \\ 
			& & (0.004) & (0.004) & (0.005) & (1.018) & (0.715) & (0.69) & (0.019) & (0.017) & (0.017) & (0.002) & (0.001) & (0.002) \\ 
			&&&&&&&&&&&&&\\
			& \multirow{2}{*}{0.60} & 0.033 & 0.024 & 0.025 & 48.268 & 32.323 & \bf 13.159 & 0.045 & 0.041 & \bf 0.031 & 1.033 & 0.987 & \bf 1.015\\ 
			& & (0.026) & (0.02) & (0.022) & (3.975) & (2.168) & (1.17) & (0.011) & (0.01) & (0.006) & (0.004) & (0.005) & (0.003) \\ 
			&&&&&&&&&&&&&\\
			& \multirow{2}{*}{0.80} & 0.258 & 0.286 & 0.261 & 1032.552 & 73.59 & \bf 20.489 & 0.384 & 0.364 &\bf  0.347 & 1.047 & 0.983 & \bf 1.013 \\ 
			& & (0.085) & (0.137) & (0.116) & (465.256) & (4.703) & (1.141) & (0.011) & (0.014) & (0.013) & (0.011) & (0.001) & (0.004) \\ 
			&&&&&&&&&&&&&\\
			\hline
			&&&&&&&&&&&&&\\
			\multirow{12}{*}{30000} & \multirow{2}{*}{0.15} & 0.002 & 0.002 & \bf 0.002 & 6.153 &  5.584 & \bf 3.036 & 0.075 & 0.075 & 0.075 & 1.014 & 0.989 & \bf 1.009 \\ 
			& & (0.001) & (0.001) & (0.001) & (0.549) & (0.363) & (0.215) & (0.004) & (0.004) & (0.004) & (0.003) & (0.001) & (0.002) \\ 
			&&&&&&&&&&&&&\\
			& \multirow{2}{*}{0.30} & 0.005 & 0.006 & 0.007 & 10.748 & 9.678 & \bf 5.334 & 0.025 & 0.025 & 0.026 & 1.014 & 0.988 & \bf 1.008 \\ 
			& & (0.003) & (0.003) & (0.003) & (1.04) & (0.896) & (0.438) & (0.006) & (0.006) & (0.006) & (0.002) & (0.001) & (0.002)  \\ 
			&&&&&&&&&&&&&\\
			& \multirow{2}{*}{0.60} & 0.023 & 0.028 & 0.037 & 26.090 & 21.774 & \bf 10.015 & 0.042 & 0.041 & \bf 0.035 & 1.022 & 0.991 & \bf 1.011 \\ 
			& & (0.017) & (0.018) & (0.021) & (2.682) & (1.792) & (0.768) & (0.011) & (0.01) & (0.009) & (0.004) & (0.001) & (0.002) \\ 
			&&&&&&&&&&&&&\\
			& \multirow{2}{*}{0.80} & 0.044 & 0.064 & 0.063 & 126.037 & 57.621 & \bf 19.923 & 0.448 & 0.427 & \bf 0.404 & 1.028 & 0.984 & \bf 1.009 \\ 
			& & (0.046) & (0.061) & (0.069) & (127.016) & (3.117) & (1.418) & (0.059) & (0.052) & (0.047) & (0.006) & (0.002) & (0.004) \\
			&&&&&&&&&&&&&\\
			\hline
			&&&&&&&&&&&&&\\ 
			\multirow{12}{*}{50000} & \multirow{2}{*}{0.15} & 0.002 & 0.003 & 0.003 & 3.189 & 3.034 & \bf 2.062 & 0.078 & 0.078 & 0.079 & 1.007 & 0.992 & \bf 1.005 \\ 
			& & (0.001) & (0.001) & (0.001) & (0.172) & (0.173) & (0.161) & (0.004) & (0.004) & (0.004) & (0.001) & (0.001) & (0.001) \\ 
			&&&&&&&&&&&&&\\ 
			& \multirow{2}{*}{0.30} & 0.002 & 0.003 & 0.003 & 5.712 & 5.370 & \bf 3.283 & 0.026 & 0.026 & 0.027 & 1.008 & 0.993 & \bf 1.004 \\ 
			& & (0.001) & (0.001) & (0.001) & (0.521) & (0.470) & (0.157) & (0.003) & (0.003) & (0.003) & (0.001) & (0.001) & (0.001) \\ 
			&&&&&&&&&&&&&\\ 
			& \multirow{2}{*}{0.60} & 0.021 & 0.020 & \bf 0.018 & 17.824 & 15.114 & \bf 6.636 & 0.050 & 0.050 & \bf 0.046 & 1.010 & 0.992 & \bf 1.006 \\ 
			& & (0.021) & (0.02) & (0.017) & (2.072) & (1.428) & (0.509) & (0.009) & (0.009) & (0.009) & (0.002) & (0.001) & (0.001) \\ 
			&&&&&&&&&&&&&\\ 
			& \multirow{2}{*}{0.80} & 0.013 & 0.019 & 0.020 & 90.804 & 29.807 & \bf 9.334 & 0.371 & 0.365 & \bf 0.352 & 1.019 & 0.991 & \bf 1.008 \\ 
			& & (0.004) & (0.010) & (0.010) & (142.017) & (4.793) & (0.800) & (0.025) & (0.024) & (0.021) & (0.003) & (0.001) & (0.002) \\ 
			&&&&&&&&&&&&&\\ 
			\hline
			\hline
		\end{tabular}
	}
\end{table}

\newpage
\begin{table}[H]
	\caption{Estimates $\widehat{\bbeta}$ for the fitted mean model. Significance is determined using $p$-values at 1 or 5\%. Significant effects are shown in bold.}\label{tab::mean-mod}
	\centering
	\resizebox*{\linewidth}{!}{
		\begin{tabular}{l|@{\extracolsep{30pt}}c|ccc|c@{}}
		\hline
		\hline
		\multirow{2}{*}{Effect} & \multirow{2}{*}{Levels} & \multirow{2}{*}{Estimates} & \multirow{2}{*}{Std. Error} & \multirow{2}{*}{Wald $z$-val} & \multirow{2}{*}{$p$-value} \\
		&&&&&\\
		\hline
		\texttt{(Intercept)} & -- &  \bf 5.3931 & 0.0368 & 146.4319 & 0.0000 \\\cline{1-2}
		\multirow{2}{*}{\texttt{Vehicle Age}} & 0--1 & \bf 0.3467 & 0.0002 & 1661.5188 & 0.0000 \\
		& 2--4 & \bf 0.2292 & 0.0002 & 1433.0911 & 0.0000 \\\cline{1-2}
		\texttt{Risk}& N & \bf 0.2679 & 0.0003 & 931.2777 & 0.0000 \\\cline{1-2}
		\multirow{3}{*}{\texttt{Age Class}} & 1 & \bf 0.5994 & 0.0021 & 286.9370 & 0.0000 \\
		& 2 & \bf 0.0480 & 0.0018 & 26.6609 & 0.0000 \\
		& 3 & \bf -0.1076 & 0.0019 & -56.9959 & 0.0000  \\\cline{1-2}
		\multirow{2}{*}{\texttt{Gender}} & F & \bf 0.0754 & 0.0041 & 18.2515 & 0.0000 \\
		& M & \bf 0.4366 & 0.0048 & 90.6983 & 0.0000 \\\cline{1-2}
		\multirow{2}{*}{\texttt{Marital Status}} & M & -1.5669 & 49.4453 & -0.0317 & 0.4874 \\
		& S & -0.1634 & 273.5849 & -0.0006 & 0.4998   \\\cline{1-2}
		\multirow{7}{*}{\texttt{Deductible Class}} & A & \bf -0.4051 & 0.0303 & -13.3525 & 0.0000 \\
		& B & \bf 0.6536 & 0.0433 & 15.1088 & 0.0000 \\
		& C & \bf -0.2605 & 0.0248 & -10.4990 & 0.0000 \\
		& D & \bf -0.2171 & 0.0247 & -8.7758 & 0.0000 \\
		& E & \bf -0.1397 & 0.0247 & -5.6566 & 0.0000 \\
		& F & \bf -0.1412 & 0.0246 & -5.7434 & 0.0000 \\
		& G & \bf -0.2576 & 0.0245 & -10.5173 & 0.0000 \\\cline{1-2}
		\multirow{4}{*}{\texttt{Marital Gender}} & A & 0.2426 & 273.5862 & 0.0009 & 0.4996 \\
		& B & 1.2976 & 49.4467 & 0.0262 & 0.4895 \\
		& C & 0.0086 & 273.5863 & 0.0000 & 0.5000  \\
		& D & 1.0249 & 49.4467 & 0.0207 & 0.4917  \\
		\hline
		\hline
	\end{tabular}
}
\end{table}
\begin{table}[H]
	\caption{Estimates $\widehat{\bgamma}$ for the fitted dispersion model. Significance is determined using $p$-values at 1 or 5\%. Significant effects are shown in bold.}\label{tab::disp-mod}
	\centering
	\resizebox*{\linewidth}{!}{
		\begin{tabular}{l|@{\extracolsep{30pt}}c|ccc|c@{}}
			\hline
			\hline
			\multirow{2}{*}{Effect} & \multirow{2}{*}{Levels} & \multirow{2}{*}{Estimates} & \multirow{2}{*}{Std. Error} & \multirow{2}{*}{Wald $z$-val} & \multirow{2}{*}{$p$-value} \\
			&&&&&\\
			\hline
			\texttt{(Intercept)} & -- & \bf 3.4421 & 0.0032 & 1080.8390 & 0.0000 \\\cline{1-2}
			\multirow{2}{*}{\texttt{Vehicle Age}} & 0--1 & \bf -0.1145 & 0.0000 & -5235.4313 & 0.0000 \\
			& 2--4 & \bf -0.0210 & 0.0000 & -1074.8869 & 0.0000 \\\cline{1-2}
			\texttt{Risk}& N & \bf 0.0425 & 0.0000 & 1428.8807 & 0.0000 \\\cline{1-2}
			\multirow{3}{*}{\texttt{Age Class}} & 1 & \bf 0.2616 & 0.0002 & 1158.0355 & 0.0000 \\
			& 2 & \bf 0.0678 & 0.0002 & 343.7858 & 0.0000 \\
			& 3 & \bf -0.0086 & 0.0002 & -40.9194 & 0.0000 \\\cline{1-2}
			\multirow{2}{*}{\texttt{Gender}}& F & \bf 0.1754 & 0.0004 & 402.4569 & 0.0000 \\
			& M & \bf 0.1334 & 0.0005 & 248.1117 & 0.0000 \\\cline{1-2}
			\multirow{2}{*}{\texttt{Marital Status}} & M & -0.5984 & 2.0075 & -0.2981 & 0.3828 \\
			& S & -1.4816 & 10.6082 & -0.1397 & 0.4445 \\\cline{1-2}
			\multirow{7}{*}{\texttt{Deductible Class}} & A & \bf -0.5718 & 0.0039 & -147.6229 & 0.0000 \\
			& B &  \bf 0.0306 & 0.0060 & 5.0807 & 0.0000 \\
			& C & \bf -0.2006 & 0.0032 & -61.9341 & 0.0000 \\
			& D & \bf 0.0162 & 0.0032 & 5.0373 & 0.0000 \\
			& E & \bf 0.0881 & 0.0032 & 27.3938 & 0.0000 \\
			& F & \bf 0.0553 & 0.0032 & 17.2634 & 0.0000 \\
			& G & \bf -0.0066 & 0.0032 & -2.0783 & 0.0188 \\\cline{1-2}
			\multirow{4}{*}{\texttt{Marital gender}} & A & 1.3861 & 10.6083 & 0.1307 & 0.4480 \\
			& B & 0.4377 & 2.0076 & 0.2180 & 0.4137 \\
			& C & 1.4736 & 10.6083 & 0.1389 & 0.4448 \\
			& D & 0.5055 & 2.0077 & 0.2518 & 0.4006 \\
			\hline
			\hline
		\end{tabular}
	}
\end{table}

\newpage

\bibliography{spatial-tweedie1}

\end{document}